\documentclass[aps,prl,reprint,superscriptaddress,floatfix]{revtex4-1}
\usepackage{graphicx}

\usepackage{amsmath}
\usepackage{amssymb}
\usepackage{hyperref}
\usepackage{bm}

\usepackage{color}

\begin{document}
\title{Switchable and unidirectional plasmonic beacons in hyperbolic 2D materials}%
\author{Andrei Nemilentsau}
\email{anemilen@umn.edu}
\affiliation{Department of Electrical \& Computer Engineering, University of Minnesota, Minneapolis, MN 55455, USA}

\author{Tobias Stauber}
\affiliation{Materials Science Factory, Instituto de Ciencia de Materiales de Madrid, CSIC, E-28049 Madrid, Spain}

\author{Guillermo G\'{o}mez-Santos}
\affiliation{Departamento de F\'{\i}sica de la Materia Condensada, Instituto Nicol\'{a}s Cabrera and Condensed Matter Physics Center (IFIMAC), Universidad Aut\'{o}noma de Madrid, E-28049 Madrid, Spain}

\author{Mitchell Luskin}
\affiliation{School of Mathematics, University of Minnesota, Minneapolis, MN 55455, USA}

\author{Tony Low}
\email{tlow@umn.edu}
\affiliation{Department of Electrical \& Computer Engineering, University of Min1nesota, Minneapolis, MN 55455, USA}

\date{\today}%

\begin{abstract}
In hyperbolic 2D materials, energy is channeled to their deep subwavelength polaritonic modes via four narrow beams. Here we consider the launching of surface polaritons in the hyperbolic 2D materials and demonstrate that efficient uni-directional excitation is possible with an elliptically polarized electric dipole, with the optimal choice of dipole ellipticity depending on the materials optical constants. The selection rules afforded by the choice of dipole polarization allow turning off up to two beams, and even three if the dipole is placed close to an edge. This makes the dipole a directionally switchable beacon for the launching of sub-difractional polaritonic beams, a potential logical gate. We develop an analytical approximation of the excitation process which describes the results of the numerical simulations well and affords a simple physical interpretation.

\end{abstract}

\maketitle
3D hyperbolic materials, i.e., strongly anisotropic materials that have metallic-type response along one of the optical axes and dielectric-type response along the other two (or vice versa), have recently attracted a lot of attention \cite{poddubny2013hyperbolic,ferrari2015hyperbolic,basov2016polaritons,low2017polaritons,smalley2018dynamically}. These materials support propagation of sub-diffractional waves over long distances, and are promising for applications such as waveguiding, \cite{caldwell2014sub,dai2015graphene,kumar2015tunable,maier2017ultracompact}, hyperlensing and focusing\cite{li2015hyperbolic,dai2015subdiffractional}, negative refraction \cite{lin2017all}, and enhancement of dipole-dipole interactions between emitters \cite{cortes2012quantum,tumkur2015control,cortes2017super}. Recently, the existence of 2D hyperbolic materials and metasurfaces, supporting in-plane hyperbolicity, was theoretically proposed \cite{gomez2015hyperbolic,nemilentsau2016anisotropic,yermakov2015hybrid,correas2015nonlocal,gangaraj2017directive} and demonstrated experimentally \cite{ma2018plane, zheng2018mid, yermakov2018experimental}. Particularly, characteristics of the surface hyperbolic polaritons in a natural vdW material, $\alpha$-MoO$_3$, have been measured \cite{ma2018plane, zheng2018mid}. Moreover, a hyperbolic metasurface was implemented in GHz frequency range using anisotropic metallic crosses printed on a dielectric substrate \cite{yermakov2018experimental}. 

\begin{figure}[h!]
	\centering
	\includegraphics[width=3in]{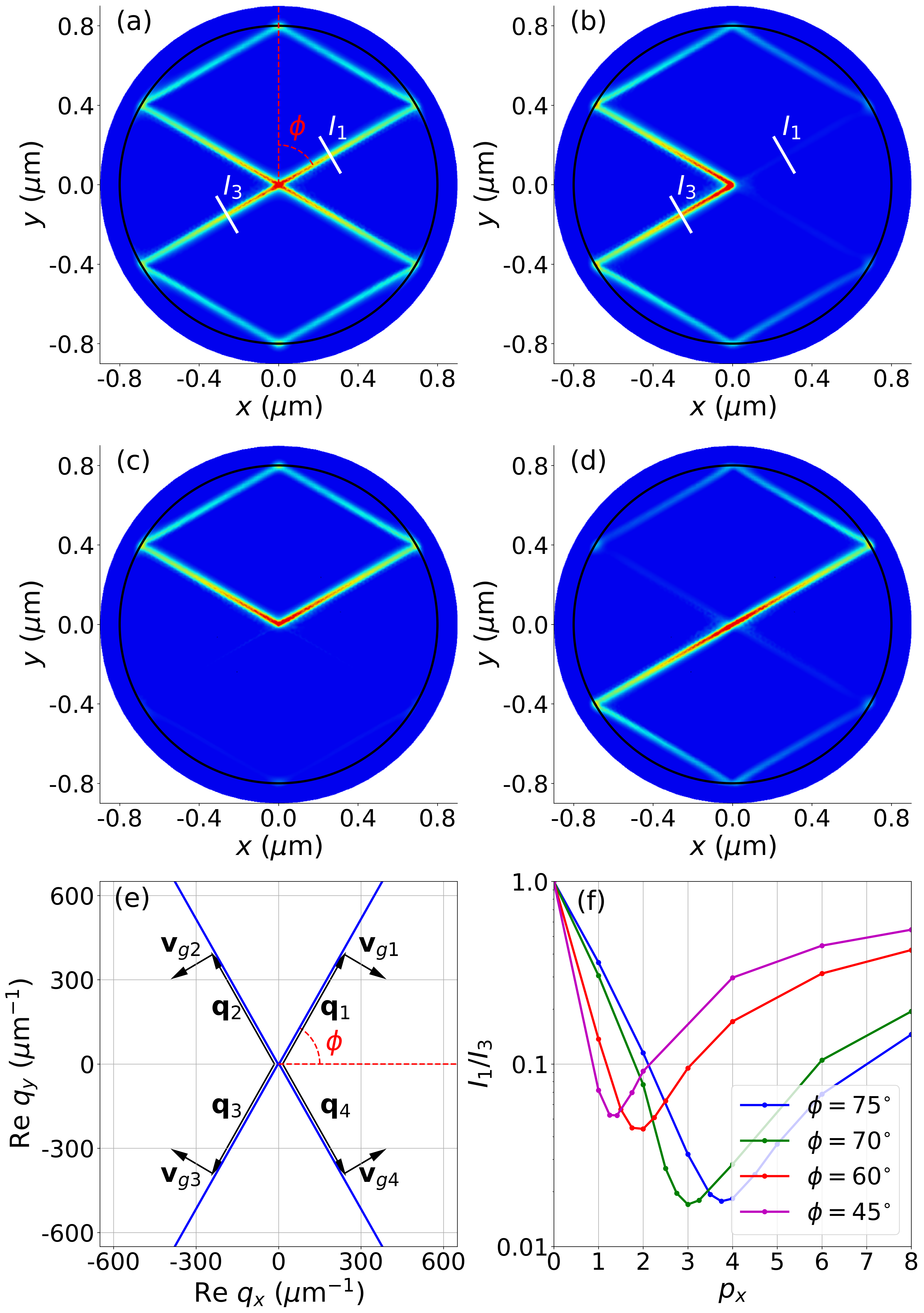}
	\caption{\textit{\textbf{Plasmon launching with elliptical dipole}}. \textbf{(a-d)}. Electric field, $\left|\mathbf{E}\right|$, of plasmons excited in a disk of hyperbolic material ($\phi = 60^{\circ}$) by an electric dipole placed in a center of a disk 5 nm above the surface. The disk radius is 800 nm. The dipole momentum is (a) $\mathbf{p} =  - i\mathbf{e}_z$ A$\cdot$m, (b) $\mathbf{p}= \mathbf{e}_x/\cos\phi - i\mathbf{e}_z = 2 \mathbf{e}_x - i\mathbf{e}_z$ A$\cdot$m, (c) $\mathbf{p} =  \mathbf{e}_y/\sin\phi - i\mathbf{e}_z= 1.15\mathbf{e}_y - i\mathbf{e}_z$ A$\cdot$m, (d) $\mathbf{p} =\mathbf{e}_x/\cos\phi - \mathbf{e}_y/\sin\phi = 2 \mathbf{e}_x - 1.15\mathbf{e}_y$ A$\cdot$m. \textbf{(e)} $k$ surface, $\omega(q_x, q_y) = \mathrm{const}$, for plasmons in the hyperbolic material in panels (a-d). \textbf{(f)} Ratio of intensities, $I_1/I_3$, carried by hyperbolic rays through detectors (white lines in panels (a-b), and Eq. \eqref{Eq:intensity}) for four different hyperbolic materials. The electric dipole momentum is $\mathbf{p} = p_x \mathbf{e}_x - i\mathbf{e}_z$ A$\cdot$m. The materials are distinguished by an angle, $\phi$, between the hyperbolic rays and the $y$-axis. $\hbar\omega = 0.19$ eV. }
	\label{fig1}
\end{figure}

2D hyperbolic materials are of particular interest as they support propagation of surface polaritons that carry energy in the form of four narrow rays \cite{gomez2015hyperbolic,nemilentsau2016anisotropic,yermakov2015hybrid}, as can be seen in Fig. \ref{fig1}a. This allows for efficient channeling of the signal from the source toward the desired target, which is crucial for the nanophotonics and applications in such fields as communication, computing, energy and quantum information. The advent of novel low loss 2D hyperbolic platform could present a paradigm shift in how energy can be steered. Typically, however, only one ray (connecting source and target) carries a signal, while the other three siphon energy away from the source. Thus it is highly desirable to develop an efficient way for the uni-directional excitation of the hyperbolic polaritons.

There are two broad approaches that can be used to mitigate the problem. The first one involves non-reciprocal materials that support uni-directional polaritonic modes that can only propagate along a given set of directions. This includes magnetoplasmons \cite{fetter1985edge,yan2012infrared,lin2013unidirectional,liu2015directional} in systems subjected to a strong static magnetic field, chiral plasmons in systems with non-zero Berry curvature \cite{kumar2016chiral,song2016chiral}, graphene sheets biased with drift electric current \cite{morgado2018drift} and topologicaly protected modes in photonic crystals with topologically non-trivial band structure \cite{lu2014topological,jin2017infrared,pan2017topologically}. Unfortunately, to implement these non-reciprocal systems proves to be quite cumbersome.

An alternative approach relies on exploiting spin-orbit interactions of light in reciprocal materials \cite{bliokh2015spin,aiello2015transverse,lodahl2017chiral}. Particularly, control over the direction of propagation of guided modes in metallic or dielectric waveguides  \cite{alonso2014controlling,evlyukhin2015resonant,sinev2017chirality,petersen2014chiral,neugebauer2014polarization} has been achieved by launching the guided modes from nano-antennas illuminated by circular polarized light fields. Moreover, the near-field of circular polarized electric and magnetic dipoles has been extensively used  for unidirectional excitation of guided modes in various isotropic reciprocal systems \cite{rodriguez2013near,le2015nanophotonic,espinosa2016transverse,picardi2017unidirectional,picardi2018janus}. Recently, 3D uniaxial hyperbolic materials were used for realizing similar spin-orbit coupling of light into polaritons \cite{kapitanova2014photonic,yermakov2016spin,jiang2018group}. Marrying the highly anisotropic optical density-of-states in a hyperbolic medium with the directional coupling via spin-orbit interaction would then enable highly efficient directional launching of surface plasmon modes.

Here, we present the theory of optimal coupling of light into polaritonic modes in hyperbolic 2D materials via general elliptical dipoles, and address the possibility of unidirectional excitation. We describe hyperbolic 2D materials as conducting sheets of zero thickness, with a conductivity tensor given by \cite{gomez2015hyperbolic,yermakov2015hybrid,nemilentsau2016anisotropic} $
\bar{\bar{\sigma}} = \textrm{diag} \{ \sigma_x, \sigma_y\} = \textrm{diag}\left\{\sigma'_x + i \sigma''_x,  \sigma'_y + i \sigma''_{y}\right\}$, where $\sigma''_{x} \sigma''_{y} < 0$, and $\sigma'$ and $\sigma''$ designate real and imaginary parts. In hyperbolic 2D materials, surface polaritons excited by a linear polarized electric dipole generally channel energy along four narrow rays as is shown in Fig. \ref{fig1}(a). We demonstrate that the efficient one-way excitation of the hyperbolic rays can be achieved by using elliptically \footnote{Elliptically polarized dipole, $\mathbf{p} = p_x \mathbf{e}_x + i p_z \mathbf{e}_z = p_+ \mathbf{p}_R + p_- \mathbf{p}_L$, is a superposition of right and left circularly polarized dipoles, $\mathbf{p}_{R,L} = \left(\mathbf{e}_x \mp i\mathbf{e}_z\right)/\sqrt{2} $, where $p_{\pm} = (p_x \mp p_z)/\sqrt{2}$.} (rather than circularly) polarized dipoles (see Fig. \ref{fig1}(b-d), where the optimum dipole polarization depends on the material conductivity. Particularly, in the case of the dipole polarized in $xz$ plane, $\mathbf{p} = p_x \mathbf{e}_x - i \mathbf{e}_z$ A$\cdot$m, the optimum dipole momentum is $p_x = \sqrt{1 + |\sigma''_x/\sigma''_y|}$ (see Fig. \ref{fig1}f). The simulations in Figs. \ref{fig1}(a-d) were performed using the Maxwell's equation solver COMSOL Multiphysics RF Module \footnote{\url{https://www.comsol.com/}}, assuming that $\sigma''_{x} = 2.85$ mS, $\sigma''_{y} = - 0.95$ mS.

In order to understand this behavior, let us consider the dispersion relation for the surface plasmons in a hyperbolic material \cite{gomez2015hyperbolic,nemilentsau2016anisotropic,yermakov2015hybrid},  $\left(q_x^2 - k_0^2\right) \sigma_{x}  + \left(q_y^2 - k_0^2\right) \sigma_{y} =   2 i \gamma_0 \omega \left(\epsilon_0 + \mu_0 \sigma_{x} \sigma_{y}/4\right)$,
where $\mathbf{q} = q_x \mathbf{e}_x + q_y \mathbf{e}_y$ is a plasmon wave vector, $\gamma_0 = \sqrt{q_x^2 + q_y^2 - k_0^2}$, $k_0 = \omega \sqrt{\epsilon_0 \mu_0}$ is the vacuum wavenumber, and $\epsilon_0$, $\mu_0$ are  vacuum permittivity and permeability, respectively. The $k$ surface (i.e., $\omega(q_x, q_y) = \mathrm{const}$) of a hyperbolic material is presented in Fig. \ref{fig1}(e). We see that the $k$ surface  takes a hyperbolic shape, with the hyperbola asymptotes making angles $\pm \phi$ with the $x$-axis, where  $\tan \phi = \sqrt{|\sigma''_{x}/\sigma''_{y}|}$ and $\phi = 60^{\circ}$. This is different from the case of an isotropic 2D material, such as graphene, where the $k$-surface is a circle.

The direction of the plasmon energy flow is defined by the group velocity, $\mathbf{v}_g = \nabla_{\mathbf{q}} \omega(\mathbf{q})$, which is orthogonal to the $k$-surface (see Fig. \ref{fig1}e). In an isotropic material, where the $k$-surface is a circle, there is no preferential direction for the normal to the $k$-surface, and thus the plasmons carry energy in all directions. In the hyperbolic material, where the $k$-surfaces are hyperbolas, the normals to the hyperbola asymptotes (and thus the group velocities) are parallel to each other. The normals in Fig. \ref{fig1}e point towards $q_x$ axis (rather than away from it) as the plasmon frequency increases along this direction\cite{luo2002all,tserkezis2010extraordinary}(see also Fig. S2 in SM \footnote{See Supplemental Material at [URL will be
	inserted by publisher] for the details on the dispersion relation of hyperbolic plasmons and derivation of the electrostatic potential induced in the hyperbolic material by an electric dipole}). Thus the hyperbolic plasmons carry energy in the form of four sub-diffractional rays (one in each of the four quadrants) in the directions making angles $\pm\phi$ with the $y$-axis (Figs. \ref{fig1}(a-d)). The points where the hyperbolic rays hit the disk edges serve as sources for the secondary hyperbolic rays.

The energy flow of the hyperbolic plasmons along the certain directions can be suppressed by choosing the polarization plane of the electric dipole (see Figs. \ref{fig1}b-d). Particularly, the elliptical dipole polarized in $xz$ plane can suppress the hyperbolic rays propagating either in first and fourth quadrants ($\mathbf{p} = \mathbf{e}_x/\cos\phi - i\mathbf{e}_z$ A$\cdot$m, Fig. \ref{fig1}b) or third and fourth quadrants ($\mathbf{p} = \mathbf{e}_y/\sin\phi - i\mathbf{e}_z$ A$\cdot$m, Fig. \ref{fig1}c). On the other hand, the linear dipole polarized in the $xy$ plane can suppress the energy flow in the second and fourth quadrants ($\mathbf{p} = \mathbf{e}_x/\cos\phi - \mathbf{e}_y/\sin\phi$ A$\cdot$m, Fig. \ref{fig1}d). Moreover, we can silence only one ray, while allowing for an excitation of the other three, if the dipole is not bound to any coordinate plane, i.e., $\mathbf{p} = 0.5(\mathbf{e}_x/\cos\phi -\mathbf{e}_y/\sin\phi) - i\mathbf{e}_z$ A$\cdot$m (see Fig. S8).  The amount of energy deposited by the dipole into each of the rays can be controlled by choosing the ellipticity of the dipole. We estimate the efficiency of the ray suppression by calculating the ratio, $I_1/I_3$, between the intensities carried by the hyperbolic rays in the first and third quadrants (see Fig. \ref{fig1}(f)), where
\begin{equation} \label{Eq:intensity}
I_i = \int_{S_i} d\mathbf{r} \left|\mathbf{E}(\mathbf{r})\right|^2.
\end{equation}
Here $S_i$ is the cross-section of the detector placed at a distance 300 nm away from the point source (see white dashes on Figs. \ref{fig1}(a-d)) and oriented orthogonally to the direction of the ray in each of the four quadrants (i.e., a normal to $S_i$ always points along the direction of the ray), $i = 1,2,3,4$.

In Fig. \ref{fig1}(f), we study the efficiency of hyperbolic ray suppression in four hyperbolic materials distinguished by an angle, $\phi$, between the rays and the $y$-axis (or between the hyperbola asymptotes to $k$-surface and the $x$-axis), i.e., $\phi = 75^{\circ}$ ($\sigma''_y = - 0.19$ mS), $\phi = 70^{\circ}$ ($\sigma''_y = -0.36$ mS), $\phi = 60^{\circ}$ ($\sigma''_y = -0.95$ mS), and $\phi = 45^{\circ}$ ($\sigma''_y = - 2.85$ mS). We considered an elliptically polarized electric dipole, $\mathbf{p} = p_x \mathbf{e}_x - i \mathbf{e}_z$ A$\cdot$m, and assumed that $\sigma''_x = 2.85$ mS. We observed efficient suppression of two out of four hyperbolic rays for an optimum value of the dipole momentum, with the intensity of the suppressed rays more than an order of magnitude weaker than that of the excited rays. We want to emphasize that the circular polarized dipole, $\mathbf{p} = \mathbf{e}_x - i \mathbf{e}_z$, does not provide efficient one-way guiding of the hyperbolic rays. Instead, the optimum value of $p_x$ depends on the material conductivity and changes between 1.44 ($\phi = 45^{\circ}$) and 4 ($\phi = 75^{\circ}$). The precise relation between $p_x$ and optical constants will be derived below.

The uni-directional excitation of the surface plasmons can be explained by studying the problem analytically in the quasi-static approximation. The electrostatic potential in the point (x,y) at the surface of the 2D material ($z=0$) induced by an electric dipole, $\mathbf{p} = (p_x, p_y, p_z)$, placed at a height $z_0$ above the 2D material, can be written as (see SI for details)
\begin{align} \label{Eq:potential}
\Phi(x,y)= -\iint \frac{dq_x dq_y}{(2\pi)^2}  & \,v_c(\mathbf{q}) t_{\mathbf{q}}  e^{i (q_x x + q_y y)} e^{ - |\mathbf{q}| |z_0|} \notag \\
&\times\left(i\mathbf{q} + |\mathbf{q}| \mathrm{sign}(z_0) \mathbf{e}_z\right) \cdot \mathbf{p},
\end{align}
where $v_c(\mathbf{q}) = 1/2\epsilon_0 |\mathbf{q}|$, $
\chi_0(\mathbf{q} )  = - (i/\omega) \mathbf{q} \cdot \underline{\sigma} \cdot \mathbf{q}$,  $t_{\mathbf{q}} = \left(1 -   v_c(\mathbf{q}) \chi_0(\mathbf{q} ) \right)^{-1}$. The electrostatic approximation, Eq. \eqref{Eq:potential}, provides a very good description of the uni-directional excitation of the plasmons in the hyperbolic 2D material (see Figs. S1-S4 for details).

Integral \eqref{Eq:potential} can be estimated analytically assuming that the dominant contribution to the integral comes from the poles in $t_{\mathbf{q}}$ (see SI). The approximate electrostatic potential is a sum of the contributions from four quadrants in the $q$-space
\begin{equation} \label{Eq:Phi_approx}
\Phi_{appr}(x,y, z=0) = \sum_{i=1}^{4} \Phi_i(x,y) \, (\mathbf{e}_i \cdot \mathbf{p}),
\end{equation}
The spatial distribution of the hyperbolic rays intensity is defined by $\Phi_i(x,y)$, where each of $\Phi_i$ is obtained by integrating \eqref{Eq:potential} over the $i$th quadrant in $q$-space, and 
\begin{equation} \label{Eq:PhiZ_approx}
\Phi_{1}(x,y) = -\theta(r_-)  \frac{q_0 e^{i q_0 (r_- + r_+)/2}}{4\epsilon_0\pi \sin 2\phi } \frac{ e^{ - q_0 \left(\tilde{z}_0 + \gamma_0 |r_-|/2\right)}  }{r_+ + i \left(2\tilde{z}_0 + \gamma_0 |r_-|\right)},
\end{equation}
$\Phi_{2}(x,y) = \Phi_{1}(-x, y)$, $\Phi_{3}(x,y) = \Phi_{1}(-x, -y) $, $\Phi_{4}(x,y)  = \Phi_{1}(x, -y)$. The strength of the dipole coupling to the hyperbolic rays is defined by the vectors $\mathbf{e}_1 = (i\cos\phi, i\sin\phi, 1)$, $\mathbf{e}_2 = (-i\cos\phi, i\sin\phi, 1)$, $\mathbf{e}_3 = (-i\cos\phi, -i\sin\phi, 1)$, $\mathbf{e}_4 = (i\cos\phi, -i\sin\phi, 1)$, which are complex conjugate of the mode vectors of the hyperbolic plasmons in each of the four quadrants (see Sec. S1 for details). Here $\theta$ is the Heaviside function, $r_{\pm} = x/\left(\sqrt{2}\sin{\phi} \right)\pm y/\left(\sqrt{2}\cos\phi \right)$, $q_0 = \sqrt{2}\epsilon_0\omega/\bar{\sigma} |\sin 2\phi|$, $ \bar{\sigma} = (\left|\sigma''_x\right| +\left|\sigma''_y\right|)/2$, $\tilde{z}_0 = z_0/(\sqrt{2}|\sin 2\phi_0|)$, $
\gamma_0 = (1/8)  \left(\sigma'_x/(\bar{\sigma}\sin^2 \phi) + \sigma'_y/(\bar{\sigma}\cos^2\phi)\right)$.
The distribution of the plasmon electrostatic potential calculated using approximate Eq. \eqref{Eq:Phi_approx} is in good agreement with the results obtained by direct numerical integration of Eq. \eqref{Eq:potential} (see Sec. S5). 

\begin{figure}[h!]
	\centering
	\includegraphics[width=3in]{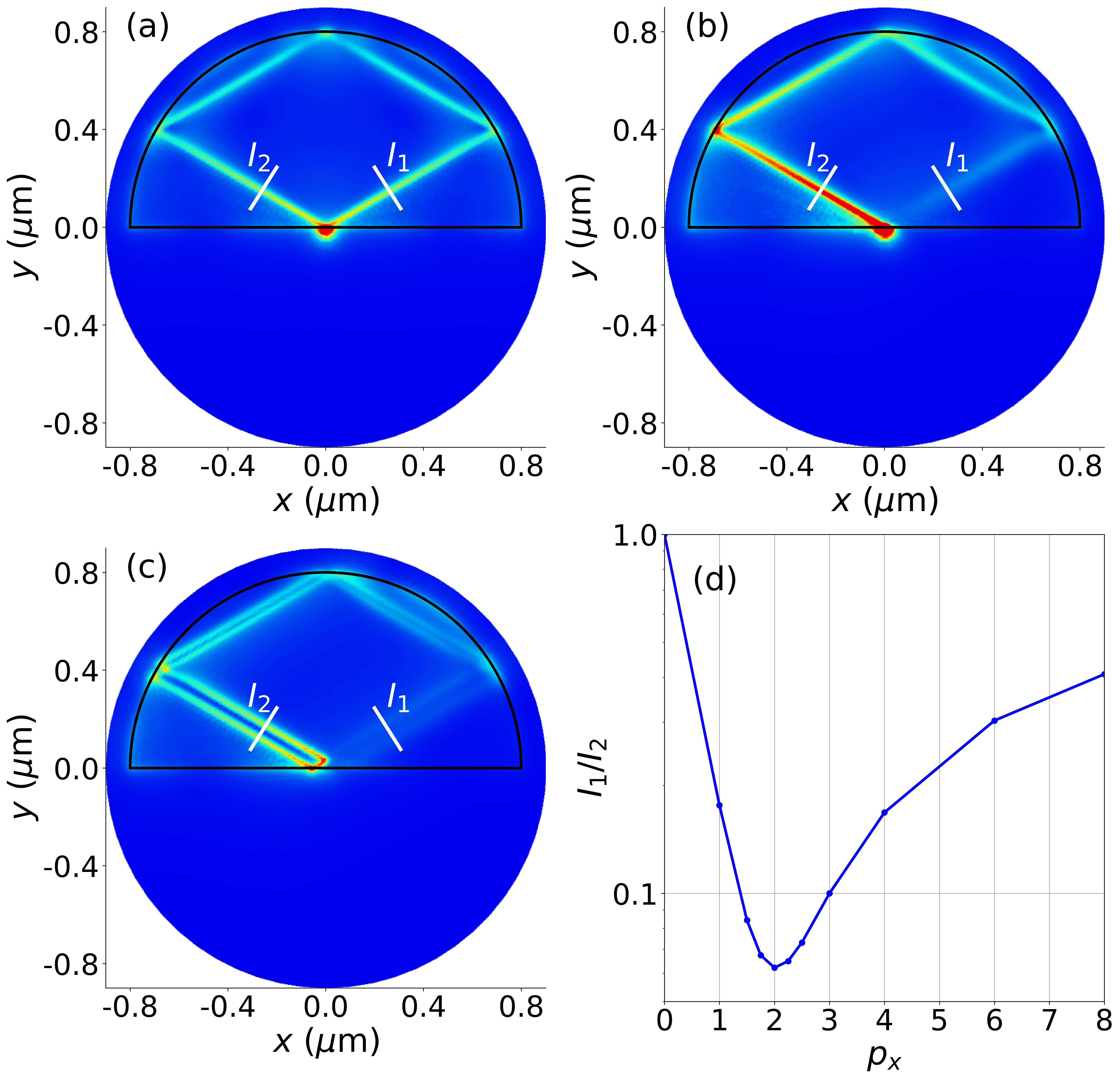}
	\caption{\textit{\textbf{Edge excitation}}. \textbf{(a-c)} Electric field, $\left|\mathbf{E}\right|$, of the plasmons excited in a half-disk of the hyperbolic material ($\phi = 60^{\circ}$) by an electric dipole, $\mathbf{p} = p_x \mathbf{e}_x-i\mathbf{e}_z$ A$\cdot$m, placed next to the disk edge ($x_0 = 0$ nm, $y_0$) 5 nm above the surface. (a) $p_x = 0$, $y_0 = 0$ nm; (b) $p_x = 2$, $y_0 = 0$ nm; (c) $p_x = 2$, $y_0 = 30$ nm. \textbf{(c)} Ratio of intensities, $I_1/I_2$, carried by plasmons in the first and second quadrants through detectors (white lines in panels (a),(b), and Eq. \eqref{Eq:intensity}).  } 
	\label{fig3}
\end{figure}

Let us consider $\Phi_1$, which is maximum when the real part of the denominator is zero, i.e., $r_+ = x/\left(\sqrt{2}\sin{\phi} \right) + y/\left(\sqrt{2}\cos\phi \right) = 0$. This defines a line, $y = - x/\tan \phi$, along the direction of hyperbolic rays in the second and fourth quadrants. The width of the line, accounting for the beam collimation, is given by the imaginary part of the denominator in Eq. \eqref{Eq:PhiZ_approx} and is mainly controlled by distance between the dipole and the 2D material. The $\Phi_{1}$ is non-zero only when $r_- = x/\left(\sqrt{2}\sin{\phi} \right) - y/\left(\sqrt{2}\cos\phi \right) >0$, or $y < x/\tan\phi $ (due to $\theta(r_-)$ factor in Eq. \eqref{Eq:PhiZ_approx}). The inequality is satisfied for the fourth quadrant and thus the term $\Phi_1$ describes the hyperbolic ray carrying energy in the fourth quadrant only. This is in agreement with the qualitative analysis presented in Fig. \ref{fig1}e, where the group velocity of the plasmons in the 1st quadrant of $q$ space points to the 4th quadrant ($\Phi_1$ originates from the integration over the first quadrant). Similarly, it is straightforward to demonstrate that $\Phi_2$ describes the hyperbolic ray in the 3rd quadrant, $\Phi_3$ --- in the 2nd quadrant, and $\Phi_4$ --- in the first quadrant.

For the dipole with momentum $\mathbf{p} = p_x \mathbf{e}_x - i \mathbf{e}_z$ A$\cdot$m, $\mathbf{e}_1 \cdot \mathbf{p} = \mathbf{e}_4 \cdot \mathbf{p} = i (p_x \cos \phi - 1) = 0$ if $p_x = 1/\cos\phi$. Thus, the dipole does not excite the hyperbolic rays $\mathbf{e}_{1,4}$ carrying energy in the 4th and 1st quadrants, respectively. On the other hand, coupling between the dipole and the rays $\mathbf{e}_{2,3}$ is maximum, $\mathbf{e}_2 \cdot \mathbf{p} = \mathbf{e}_3 \cdot \mathbf{p} =- i (p_x \cos \phi + 1) = -2 i$. Thus, the dipole excites only the hyperbolic rays propagating in the second and third quadrants, while suppressing the hyperbolic rays propagating in the first and fourth quadrants. Moreover, the actual value of the dipole momentum that allows maximum suppression depends on the material conductivity through the angle, $\phi$, as
\begin{equation} \label{Eq:dipole_cond}
p_x = 1/\cos\phi = \sqrt{1 + |\sigma''_x/\sigma''_y|}.
\end{equation}
In particular, $p_x = 4, 3, 2, \sqrt{2}$ when $\theta = 15^{\circ}$, 20$^{\circ}$, 30$^{\circ}$, 45$^{\circ}$. In summary, Eq. \eqref{Eq:Phi_approx} shows that each beam can be individually addressed and even silenced by means of the selection rule encoded in the term, $\mathbf{e}_i \cdot \mathbf{p}$ (see Sec. S5.9). These results are in a good agreement with the simulations results presented in Fig. \ref{fig1}.

We want to stress that an electric dipole can not launch a single hyperbolic ray while suppressing the other three. A truly uni-directional excitation of a single hyperbolic ray can be achieved by placing an electric dipole at the edge of the material, as is shown in Figs. \ref{fig3}a,b. Particularly, by using the dipole $\mathbf{p} = 2 \mathbf{e}_x-i\mathbf{e}_z$ A$\cdot$m placed at the hyperbolic material edge it is possible to excite only one hyperbolic ray (see Fig. \ref{fig3}(b)). Moreover, the intensity of the suppressed hyperbolic ray is more than an order of magnitude lower than the intensity of the excited hyperbolic ray (see Fig. \ref{fig3}(d)). Also, hypothetically, one could switch off the third beam by using magnetic dipole moment (albeit gigantic)\cite{picardi2018janus}, in a combination with an electric dipole moment. Excitation of a single beam is also possible in a disk of an anisotropic material as is discussed in Sec. S7.

In order to get more insight in the process of launching of the hyperbolic mode from the edge, we considered the elliptically polarized dipole placed 30 nm away from the half-disk edge (see Fig. \ref{fig3}c). The dipole still launches two hyperbolic rays, as is the case for a full disk (see Fig. \ref{fig1}b). However, the second ray is back reflected from the edge and thus both of the rays carry energy in the same direction. The total intensity is additive when the two rays can be clearly resolved, as in Fig. \ref{fig3}c, so that the energy carried by these two rays through the detector is 	approximately twice the energy of each of the single rays in \ref{fig1}b. When we move the dipole closer to the disk edge (Fig. \ref{fig3}b), the two rays merge and interfere constructively, as inferred from the fact that the energy flow increases almost four fold compared to that of each of the single rays in Fig. \ref{fig1}b.

This constructive interference can be understood using the following simple model. The effect of the edge can be approximated by placing an additional fictitious dipole at the usual image position. The fictitious dipole polarization should be chosen to enforce zero normal component of the total field at the edge, so that the normal current also vanishes. For the case considered in Figs. \ref{fig3}b,c, this condition prescribes that the dipole and its image should be identical. Therefore, when the two dipoles merge, the total dipole doubles, and the intensity of the only surviving ray (in the physical region) quadruples. The further control of the energy flow is possible by placing the dipole near the edges of more complicated shape (see Sec. S6).

\begin{figure}[h!]
	\centering
	\includegraphics[width=3in]{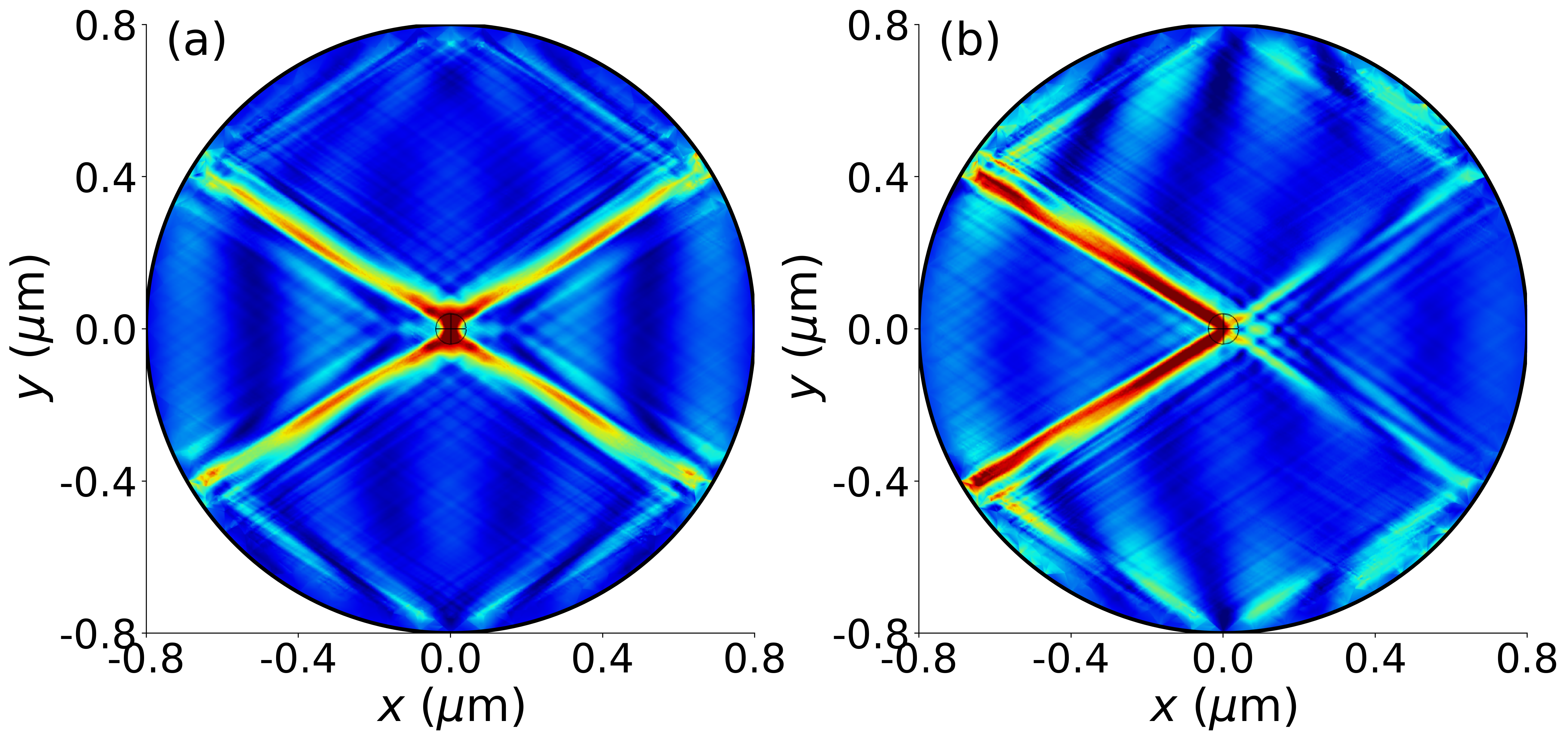}
	\caption{\textit{\textbf{Plane wave excitations}}. Unidirectional excitation of plasmons in a hyperbolic material ($\phi = 60^{\circ}$) by an elliptically polarized plane wave, $\mathbf{E} = \mathbf{e}_p e^{i k_y y}$. A metallic sphere of radius 40 nm and relative permittivity, $\epsilon_m = -2$, is placed on top of a 2D material. (a) $\mathbf{e}_p = -i\mathbf{e}_z$, (b)  $\mathbf{e}_p = 2\mathbf{e}_x - i \mathbf{e}_z$. }
	\label{fig4}
\end{figure}

Finally, let us consider an experimental possibility of the uni-directional launching of the hyperbolic rays. In order to do this we place a metallic sphere of radius 40 nm and relative permittivity $\epsilon_m = -2$ on top of the hyperbolic material. The hyperbolic plasmon is then launched by illuminating the system with a plane electromagnetic wave propagating along $y$ direction, $\mathbf{E} = \mathbf{e}_p e^{i k_y y}$. The electromagnetic wave excites plasmons in the metallic sphere, which acts now as an effective electric dipole and thus can effectively couple to the plasmons in the hyperbolic material. Indeed, as can be seen in Fig. \ref{fig4}a, the linearly polarized plane wave, $\mathbf{e}_p = - i \mathbf{e}_z$, excites all four hyperbolic rays with equal efficiency. However, by using an elliptically polarized wave, $\mathbf{e}_p = 2 \mathbf{e}_x - i \mathbf{e}_z$, two out of four hyperbolic rays can be efficiently suppressed (see Fig. \ref{fig4}(b)). Recent development of resonant metal antennas for 2D plasmonics suggests  such experimental setup is feasible \cite{alonso2014controlling}.

Concluding, we studied switchable plasmonic beacons and unidirectional excitation of the surface plasmons in a hyperbolic 2D material. We demonstrated that efficient unidirectional launching of hyperbolic rays requires an elliptically polarized electric dipole rather than a circular polarized one. Moreover, the dipole ellipticity depends on the direction of the hyperbolic rays propagation, i.e. on the material conductivity. In general, we can only suppress two out of four hyperbolic rays by using an electric dipole. However, we can excite a single hyperbolic ray by launching plasmons at the edge of the hyperbolic material. The coherent interference which lies at the heart of this work does not have to be confined to different component of a single dipole. One can easily envision the manifold of possibilities that the presence of two or more dipoles will open, potentially making 2D hyperbolic materials an ideal platform for polaritonic beam steering.

AN, TL, and ML have been partially supported by the Army Research Office (ARO) Multidisciplinary University Research Initiative (MURI) Award No. W911NF-14-1-0247. TL and AN acknowledges partial support from NSF/EFRI- 1741660. AN, TS, ML, and TL acknowledge the hospitality of the Institute for Mathematics and its Applications. TS and GS acknoledge support from Spain’s MINECO under Grants No. MDM-2014-0377, No. FIS2017-82260-P, and No. FIS2015-64886-C5-5-P.

\newpage
\begin{widetext}
{\bf \huge Supplementary Material}

\section{{Dispersion and mode structure of the hyperbolic plasmons}}

{We consider an anisotropic two-dimensional material
\begin{equation}
\underline{\sigma} = \left(
\begin{array}{cc}
\sigma_{x} & 0 \\
0 & \sigma_{y}
\end{array}
\right).
\end{equation}
We choose the coordinate system rotated counterclockwise by an angle $\varphi$ with respect to the coordinate system aligned with the material optical axes, so that the plasmon propagation direction always coincides with the $x$ axis of the rotated coordinate system, i.e.
\begin{equation}
\mathbf{k} = q_x \mathbf{e}_x + q_y \mathbf{e}_y \pm i\mathbf{e}_z \gamma_0 = q \mathbf{e}'_x \pm i\mathbf{e}_z \gamma_0,
\end{equation}
where $q = \sqrt{q_x^2 + q_y^2}$,  $\gamma_0^2 = q^2 - k_0^2$, $k_0^2 = \omega^2 \mu_0 \varepsilon_0$ and $\mathbf{e}'_x$ is a unit vector along the $x$ axis of the rotated system. In the rotated coordinate system the conductivity tensor takes form
\begin{equation}
\underline{\sigma} =\left(
\begin{array}{cc}
\tilde{\sigma}_{xx} & \tilde{\sigma}_{xy} \\
\tilde{\sigma}_{xy} & \tilde{\sigma}_{yy}
\end{array}
\right) = \left(
\begin{array}{cc}
\sigma_x \cos^2 \varphi + \sigma_y \sin^2\varphi  & (\sigma_y - \sigma_x) \sin\varphi \cos\varphi \\
(\sigma_y - \sigma_x) \sin\varphi \cos\varphi & \sigma_x \sin^2 \varphi + \sigma_y \cos^2\varphi
\end{array}
\right).
\end{equation}}

{
EM field of the plasmon can be presented as a superposition of TE and TM modes, i.e. $\mathbf{E} = \mathbf{E}_{TE} + \mathbf{E}_{TM}$, $\mathbf{H} = \mathbf{H}_{TE} + \mathbf{H}_{TM}$, where
\begin{align} 
&\mathbf{E}_{TE} = \mathbf{e}'_y E_0 e^{i \mathbf{k}\cdot \mathbf{r}}, \quad \mathbf{H}_{TE} =  \frac{1}{i\omega \mu_0} \, \nabla \times \mathbf{E}_{TE} =  \frac{\left( \mathbf{e}'_x k_{\rho} \pm i\mathbf{e}_z \gamma_0\right) \times \mathbf{e}'_y E_0}{\omega\mu_0 }, \label{Eq:TE}\\
&\mathbf{H}_{TM} =\mathbf{e}'_y  H_0 e^{i \mathbf{k}\cdot \mathbf{r}}, \quad \mathbf{E}_{TM} = -\frac{1}{i\omega \varepsilon_0} \, \nabla \times \mathbf{H}_{TM} =  - \frac{\left( \mathbf{e}'_x k_{\rho} \pm i\mathbf{e}_z \gamma_0\right) \times \mathbf{e}'_y  H_0}{\omega\varepsilon_0},  \label{Eq:TM}
\end{align}
where in $\pm$ the '+' sign corresponds to the field above the 2D material (i.e. $z > 0$), while the '-' sign corresponds to the field below the 2D material (i.e. $z < 0$). Here we took into account that for the TE mode $\mathbf{E} \cdot \mathbf{e}'_x = 0$, while for the TM mode  $\mathbf{H} \cdot \mathbf{e}'_x = 0$. Finally we need to impose boundary conditions for the electromagnetic field across the 2D material,
\begin{align}
&\mathbf{e}_z \times \left(\mathbf{E}|_{z = 0^+} - \mathbf{E}|_{z = 0^-} \right) =  0, \qquad \mathbf{e}_z \times \left(\mathbf{H}|_{z = 0^+} - \mathbf{H}|_{z = 0^-} \right) =  \underline{\sigma} \cdot \mathbf{E}|_{z = 0}. \label{Eq:boundary_cond}
\end{align}
This leads to the plasmon dispersion in the rotated coordinate system
\begin{align} \label{Eq:plasmon}
\left( 2 +  \frac{i \gamma_0}{\omega\varepsilon_0}  \tilde{\sigma}_{xx}  \right)  \left(\frac{2 i \gamma_0}{\omega\mu_0} + \tilde{\sigma}_{yy} \right) =  \frac{i\gamma_0}{\omega\varepsilon_0}   \tilde{\sigma}_{xy}^2,
\end{align}
which can be reduced to that in the coordinate system aligned with the optical axes of the anisotropic 2D material, i.e.
\begin{align} \label{Eq:plasmon1}
2 i \gamma_0 \omega \left(\frac{ \mu_0 \sigma_{x} \sigma_{y}}{4 }  +\varepsilon_0 \right)     = \left( q_y^2  - k_0^2 \right) \sigma_{y}  + \left( q_x^2  - k_0^2 \right) \sigma_{x}.  
\end{align}}

\section{Dispersion of a quasi-static plasmon}
Let us consider 2D material in the plane $z = 0$. The charge density and current of the surface plasmon are distributed in the 2D plane as
\begin{equation}
\rho(\boldsymbol{r}_{\parallel}) = \rho(\mathbf{q}) e^{i \mathbf{q}\cdot \boldsymbol{r}_{\parallel}}, \qquad \mathbf{j}(\boldsymbol{r}_{\parallel}) = \mathbf{j}(\mathbf{q}) e^{i \mathbf{q}\cdot \boldsymbol{r}_{\parallel}},
\end{equation}
where $\mathbf{q} = q_x \mathbf{e}_x + q_y \mathbf{e}_y$ is the 2D wavevector, and $\boldsymbol{r}_{\parallel}$ is a 2D component of the radius-vector. From the continuity equation it follows that
\begin{align*}
\frac{\partial\rho}{\partial t}& = - \nabla_{\boldsymbol{r}_{\parallel}} \cdot \mathbf{j}  \quad \Rightarrow \quad  \omega \rho(\boldsymbol{r}_{\parallel}) e^{i \mathbf{q}\cdot \boldsymbol{r}_{\parallel}} = \mathbf{q} \cdot \mathbf{j}(\mathbf{q}) e^{i \mathbf{q}\cdot \boldsymbol{r}_{\parallel}} \\
& = \mathbf{q} \cdot \underline{\sigma} \cdot \mathbf{E}(\boldsymbol{r}_{\parallel}, z = 0) = - \mathbf{q} \cdot \underline{\sigma} \cdot \nabla_{\boldsymbol{r}_{\parallel}} {\Phi}(\boldsymbol{r}_{\parallel}, z = 0) = - i \Phi(\mathbf{q}) e^{i \mathbf{q}\cdot \boldsymbol{r}_{\parallel}} \, \mathbf{q} \cdot \underline{\sigma} \cdot \mathbf{q}.
\end{align*}
Thus
\begin{equation}
\omega \rho(\boldsymbol{r}_{\parallel}) = - i \Phi(\mathbf{q} ) \,\mathbf{q} \cdot \underline{\sigma} \cdot \mathbf{q},
\end{equation}
where the electrostatic potential of the plasmon in the 2D material is defined as
\begin{equation}
\Phi(\boldsymbol{r}_{\parallel}, z = 0) = \Phi(\mathbf{q}, z = 0)  e^{i \mathbf{q}\cdot \boldsymbol{r}_{\parallel}} = \Phi(\mathbf{q})  e^{i \mathbf{q}\cdot \boldsymbol{r}_{\parallel}},
\end{equation}
and $\underline{\sigma}$ is the 2D material conductivity tensor. If we define charge-charge response function as
\begin{equation}
\rho(\mathbf{q} )  = \chi_0(\mathbf{q} )  \Phi(\mathbf{q} ) ,
\end{equation}
we obtain
\begin{equation}
\chi_0(\mathbf{q} )  = - \frac{i}{\omega } \, \mathbf{q} \cdot \underline{\sigma} \cdot \mathbf{q}.
\end{equation}

The total potential is a superposition of the external and induced potentials, $\Phi = \Phi_{ext} + \Phi_{ind}$. We can define charge-charge response function with respect to the total potential as follows
\begin{equation}
\rho(\mathbf{q} )  = \chi_0(\mathbf{q} )  \Phi(\mathbf{q} )  = \chi(\mathbf{q} )  \Phi_{ext}(\mathbf{q} ) .
\end{equation}
In order to find the induced potential we use the Poisson equation,
\begin{equation} \label{Eq:Poisson}
\nabla^2 \Phi_{ind}(\mathbf{r}) =  - \frac{\delta(z)  \rho(\boldsymbol{r}_{\parallel}) }{\varepsilon_0}.
\end{equation}
Taking the Fourier transform of Eq. \eqref{Eq:Poisson}, we obtain
\begin{align}\label{Eq:Poisson_tran}
\left(|\mathbf{q}|^2 + k
_z^2 \right) \Phi_{ind}(\mathbf{q}, k_z) = \frac{  \rho(\mathbf{q})}{\epsilon_0} =  \frac{  \chi_0(\mathbf{q}) \Phi(\mathbf{q}, z = 0)}{\epsilon_0}.
\end{align}
Considering that
\begin{align*}
\Phi_{ind}(\mathbf{q}, z)  =  \int_{-\infty}^{\infty} \frac{dk_z }{2\pi} e^{ i k_z z} \Phi_{ind}(\mathbf{q}, k_z)  \quad \Rightarrow \quad \Phi_{ind}(\mathbf{q},z = 0) =  \int_{-\infty}^{\infty} \frac{dk_z }{2\pi} \Phi_{ind}(\mathbf{q}, k_z),
\end{align*}
we can further transform Eq. \eqref{Eq:Poisson_tran},
\begin{align*}
\int_{-\infty}^{\infty}\frac{d k_z}{2\pi} \, \Phi_{ind}(\mathbf{q}, k_z) =  \int_{-\infty}^{\infty}\frac{d k_z}{2\pi \left(|\mathbf{q}|^2 + k_z^2 \right)} \frac{\chi_0(\mathbf{q}) \, \Phi(\mathbf{q}, z = 0)}{\epsilon_0},
\end{align*}
\begin{align*}
\Phi_{ind}(\mathbf{q},z = 0) = \frac{\chi_0(\mathbf{q})  \Phi(\mathbf{q}, z = 0)}{2 |\mathbf{q}| \epsilon_0}.
\end{align*}
Thus,
\begin{align*}
\chi_0(\mathbf{q}) \Phi(\mathbf{q},z=0) = \chi(\mathbf{q}) \left(\Phi(\mathbf{q},z=0) - \Phi_{ind}(\mathbf{q},z=0\right) = \chi(\mathbf{q}) \left(\Phi(\mathbf{q},z=0) - \frac{\chi_0(\mathbf{q}) \Phi(\mathbf{q},z=0)}{2 |\mathbf{q}|\epsilon_0 }\right).
\end{align*}
This means that
\begin{equation} \label{Eq:charge-charge}
\chi(\mathbf{q}) = \chi_0(\mathbf{q}) \left(1 - v_c(\mathbf{q}) \chi_0(\mathbf{q}) \right)^{-1},
\end{equation}
where $v_c(\mathbf{q}) = 1/2\epsilon_0 |\mathbf{q}|$.

The dispersion of the plasmons is defined as a zero of denominator of Eq. \eqref{Eq:charge-charge}, i.e.,
\begin{equation}
1 + \frac{i}{2 \omega \epsilon_0 |\mathbf{q}|}  \, \mathbf{q} \cdot \underline{\sigma} \cdot \mathbf{q} = 0.
\end{equation}
Assuming that the conductivity is diagonal, we obtain
\begin{equation}
1 + \frac{i}{2 \omega \epsilon_0 |\mathbf{q}|}  \, \left(q_x^2 \sigma_x  + q_y^2 \sigma_y \right) = 0.
\end{equation}
Finally, neglecting losses, the dispersion of plasmons takes form,
\begin{equation} \label{Eq:dispersion_quas}
1 - \frac{1}{2\omega\epsilon_0 |\mathbf{q}|} \, \left(q_x^2 \sigma''_x + q_y^2 \sigma''_y\right) = 0,
\end{equation}
where $\sigma_x = i \sigma''_x$, $\sigma_y = i \sigma''_y$.

In the next section, we will also need the relation between total electric potential and external potential, i.e.,
\begin{equation}
\frac{\Phi(\mathbf{q},z=0)}{\Phi_{ext}(\mathbf{q},z=0)} = t_{\mathbf{q}}.
\end{equation}
Let us take into account that $\Phi_{ext}(\mathbf{q},z=0) = \Phi (\mathbf{q},z=0)- \Phi_{ind} (\mathbf{q},z=0)= \Phi(\mathbf{q},z=0) \left(1 - v_c(\mathbf{q}) \chi_0(\mathbf{q})\right)$.
Thus,
\begin{equation}
\frac{\Phi(\mathbf{q},z=0)}{\Phi_{ext}(\mathbf{q},z=0)} = t_{\mathbf{q}} = \frac{1}{1 - v_c(\mathbf{q}) \chi_0(\mathbf{q})} .
\end{equation}

\section{Dipole electric field}
In order to calculate the electrostatic potential of an electric dipole, let us consider the Green's function for the Poisson equation first,
\begin{equation}
\Delta G_{ext}(\mathbf{r}, \mathbf{r}_0) = - \delta(\mathbf{r} - \mathbf{r}_0),
\end{equation}
where $\mathbf{r} = (x, y, z)$, while $\mathbf{r}_0 = (0, 0, z_0)$. Thus
\begin{equation}
\Delta G_{ext}(\mathbf{r}, \mathbf{r}_0) = - \delta(x) \delta(y) \delta(z-z_0).
\end{equation}
We are looking for the solution in the form
\begin{equation}
G_{ext} (\mathbf{r}, \mathbf{r}_0) = \iint \frac{dq_x dq_y}{(2\pi)^2} e^{i (q_x x + q_y y)} G(z, z_0)
\end{equation}
and take into account that
\begin{equation}
\delta(x) = \int \frac{dq_x \, e^{iq_x x }}{2\pi}.
\end{equation}
Thus we obtain
\begin{equation}
\left(\frac{d^2 }{d_z^2} - q_x^2 - q_y^2\right)G(z, z_0) = -\delta(z-z_0).
\end{equation}
The solution of the above equation is
\begin{equation}
G(z,z_0) = \frac{e^{- |\mathbf{q}| |z - z_0| }}{2 |\mathbf{q}|}
\end{equation}
and the Green's function takes form
\begin{equation}
G_{ext} (\mathbf{r}, \mathbf{r}_0) =  \iint \frac{dq_x dq_y}{(2\pi)^2} \frac{ e^{i (q_x x + q_y y)} e^{ - |\mathbf{q}| |z - z_0|} }{2 |\mathbf{q}|}.
\end{equation}
{
Let us take into account that the potential in the point $\mathbf{r}$ induced by an electric dipole located in the point $\mathbf{r}_0$}, takes form
\begin{align*}
\Phi_{ext}(\mathbf{r}) & =  \frac{ \mathbf{p} \cdot (\mathbf{r}- \mathbf{r}_0)}{4\pi\epsilon_0 |\mathbf{r}- \mathbf{r}_0|^3} = - \mathbf{p} \cdot \nabla_{\mathbf{r}} \frac{ 1 }{4\pi\epsilon_0 |\mathbf{r}- \mathbf{r}_0|}.
\end{align*}
Then
\begin{align}
\Phi_{ext}(\mathbf{r}) & = -\mathbf{p} \cdot \nabla \frac{G_{ext} (\mathbf{r}, \mathbf{r}_0)}{\epsilon_0} = -\mathbf{p} \cdot \nabla  \iint \frac{dq_x dq_y}{(2\pi)^2} \,v_c(\mathbf{q})  e^{i (q_x x + q_y y)} e^{ - |\mathbf{q}| |z - z_0|} \notag \\
& = -\iint \frac{dq_x dq_y}{(2\pi)^2} \,v_c(\mathbf{q})  e^{i (q_x x + q_y y)} e^{ - |\mathbf{q}| |z - z_0|} \left(i\mathbf{q}  - |\mathbf{q}| \mathrm{sign}(z-z_0) \mathbf{e}_z\right) \cdot \mathbf{p}. \label{Eq:Phi_ext}
\end{align}
{As the external potential in the point $(x, y, z = 0)$ at the 2D material surface is}
\begin{align}
\Phi_{ext}(x,y, z = 0) & = -\iint \frac{dq_x dq_y}{(2\pi)^2} \,v_c(\mathbf{q})  e^{i (q_x x + q_y y)} e^{ - |\mathbf{q}| |z_0|} \left(i\mathbf{q} + |\mathbf{q}| \mathrm{sign}(z_0) \mathbf{e}_z\right) \cdot \mathbf{p},
\end{align}
{the total potential takes form}
\begin{align} \label{Eq:phi_tot_2D}
\Phi(x,y, z = 0 ) & = -\iint \frac{dq_x dq_y}{(2\pi)^2} \,v_c(\mathbf{q}) t_{\mathbf{q}}  e^{i (q_x x + q_y y)} e^{ - |\mathbf{q}| |z_0|} \left(i\mathbf{q} + |\mathbf{q}| \mathrm{sign}(z_0) \mathbf{e}_z\right) \cdot \mathbf{p}
\end{align}

In order to calculate the potential for $z\neq0$, we take into account that
\begin{align*}
\Phi_{ind}(\mathbf{q}, z)  =  \int_{-\infty}^{\infty} \frac{dk_z }{2\pi } e^{ i k_z z} \Phi_{ind}(\mathbf{q}, k_z) = \frac{\chi_0(\mathbf{q}) \, \Phi(\mathbf{q}, z = 0)}{2\pi\epsilon_0} \int_{-\infty}^{\infty} \frac{dk_z \, e^{ i k_z z} }{ |\mathbf{q}|^2 + k_z^2 }.
\end{align*}
The integral over $k_z$ is equal to
\begin{equation*}
\int_{-\infty}^{\infty} \frac{dk_z}{k_z - i |\mathbf{q}|}\frac{ e^{ i k_z z} }{ k_z + i |\mathbf{q}| } + \int_{\mathrm{Im} k_z > 0, |k_z|  = \infty} dk_z \cdots = 2\pi i \frac{e^{-|\mathbf{q}|z}}{2 i |\mathbf{q}|}= \pi \frac{e^{-|\mathbf{q}|z}}{|\mathbf{q}|}
\end{equation*}
if $z > 0$, and
\begin{equation*}
\int_{-\infty}^{\infty} \frac{dk_z}{k_z + i |\mathbf{q}|}\frac{dk_z e^{ -i k_z |z|} }{ k_z - i |\mathbf{q}| } + \int_{\mathrm{Im} k_z < 0, |k_z|  = \infty} dk_z \cdots = - 2\pi i \frac{e^{-|\mathbf{q}||z|}}{-2 i |\mathbf{q}|} = \pi \frac{e^{-|\mathbf{q}||z|}}{|\mathbf{q}|}
\end{equation*}
if $z < 0$ (we have minus in front of 2$\pi i$ as the contour integral is taken clockwise).
Thus
\begin{equation}
\Phi_{ind}(\mathbf{q}, z) = \frac{ e^{-|\mathbf{q}||z|}  \chi_0(\mathbf{q}) \, \Phi(\mathbf{q}, z = 0)}{2 |\mathbf{q}|\epsilon_0} = e^{-|\mathbf{q}||z|} \Phi_{ind}(\mathbf{q},z = 0).
\end{equation}
Let us take into account that
\begin{equation*}
\Phi_{ind}(\mathbf{q},z = 0) = v_c(\mathbf{q}) \chi_0(\mathbf{q}) \Phi(\mathbf{q},z = 0) = v_c(\mathbf{q}) \chi_0(\mathbf{q}) \left(\Phi_{ind}(\mathbf{q},z = 0) + \Phi_{ext}(\mathbf{q},z = 0) \right),
\end{equation*}
or
\begin{equation*}
\Phi_{ind}(\mathbf{q},z = 0) = t_{\mathbf{q}} v_c(\mathbf{q}) \chi_0(\mathbf{q}) \Phi_{ext}(\mathbf{q},z = 0).
\end{equation*}
Finally,
\begin{equation}
\Phi_{ind}(\mathbf{q}, z) = e^{-|\mathbf{q}|z} t_{\mathbf{q}} v_c(\mathbf{q}) \chi_0(\mathbf{q}) \Phi_{ext}(\mathbf{q},z = 0).
\end{equation}
Thus
\begin{align}
\Phi_{ind}(\mathbf{r}) & = -\iint \frac{dq_x dq_y}{(2\pi)^2} \,v_c^2 (\mathbf{q}) \chi_0(\mathbf{q}) t_{\mathbf{q}}  e^{i (q_x x + q_y y)} e^{ - |\mathbf{q}| (|z_0| + |z|)}  \left(i\mathbf{q} +  \mathrm{sign}(z_0) |\mathbf{q}| \mathbf{e}_z\right) \cdot \mathbf{p}.
\end{align}
{And the total potential is defined by
\begin{equation}\label{Eq:Phi_tot_above}
\Phi(\mathbf{r}) = \Phi_{ind}(\mathbf{r}) + \Phi_{ext}(\mathbf{r}),
\end{equation}
where  $\Phi_{ext}(\mathbf{r})$ is given by \eqref{Eq:Phi_ext}.}

\section{Hyperbolic material}
Let us assume that $\sigma_x = \sigma'_x + i |\sigma''_x|$, $\sigma_y = \sigma'_y - i |\sigma''_y|$. The iso-frequency surfaces in a hyperbolic material are shown in Fig. \ref{Fig:dispersion}.  Asymptotics to the iso-frequency surfaces in a hyperbolic material are defined by an angle
\begin{equation} \label{Eq:asymptote}
\tan \phi = \pm \sqrt{\left|\frac{\sigma''_x}{\sigma''_y}\right|}.
\end{equation}

\begin{figure}[h!]
	\centering
	\includegraphics[width=5in]{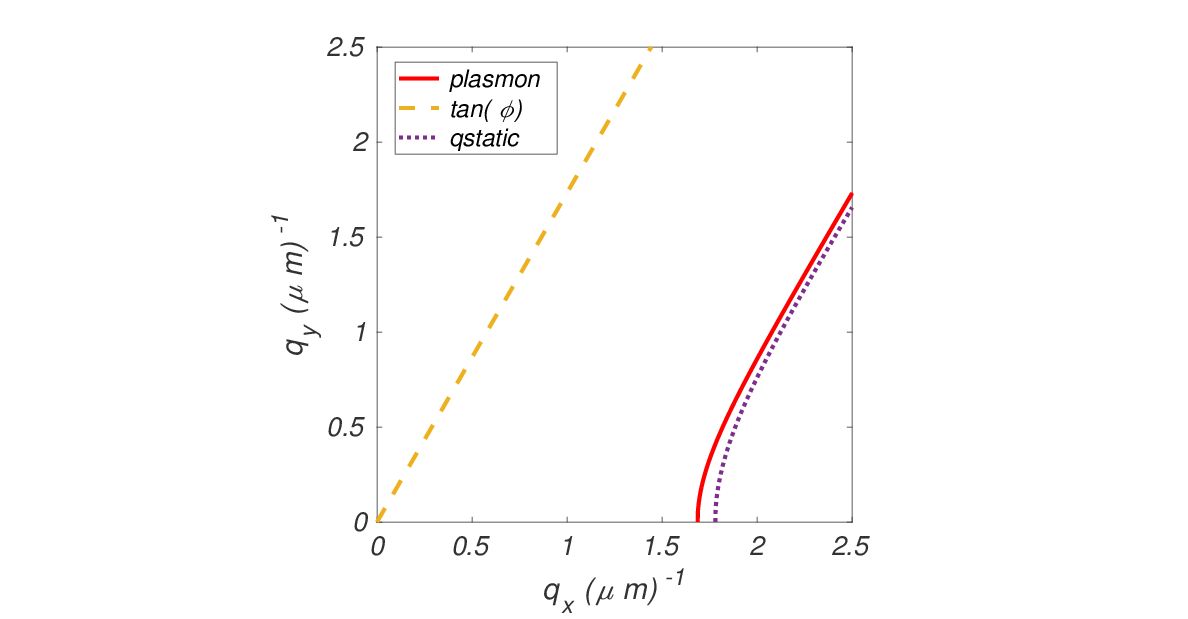}
	\caption{Iso-frequency surfaces for surface plasmons in a hyperbolic material. Calculations were made accounting for (red line, Eq. \eqref{Eq:plasmon1}) and excluding the retardation effects (dotted violet line, Eq. \eqref{Eq:dispersion_quas}). $\sigma''_{x} = 2.85$ mS, $\sigma''_{y} = - 0.95$ mS, $\phi = 60^{\circ}$, $\hbar\omega = 0.19$ eV.  } \label{Fig:dispersion}
\end{figure}

\begin{figure}[h!]
	\centering
	\includegraphics[width=5in]{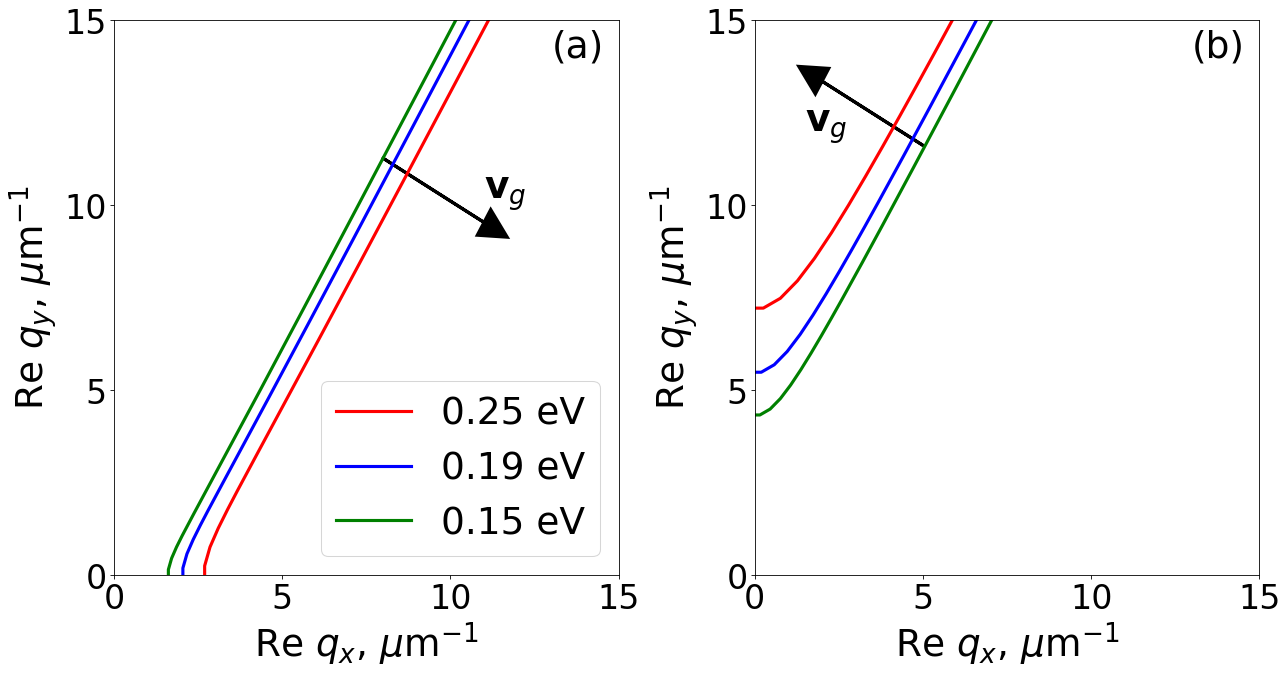}
	\caption{{Group velocity, $\mathbf{v}_g = \nabla_{\mathbf{q}} \omega(\mathbf{q})$, for the plasmons in the hyperbolic material, calculated using Eq. \eqref{Eq:plasmon1}. (a) $\sigma''_{x} = 2.85$ mS, $\sigma''_{y} = - 0.95$ mS. (b) $\sigma''_{x} = -2.85$ mS, $\sigma''_{y} =  0.95$ mS. } } \label{Fig:group_vel}
\end{figure}

{In the anisotropic material the direction of the energy flow coincides with the group velocity, $\mathbf{v}_g = \nabla_{\mathbf{q}} \omega(\mathbf{q})$. The direction where the group velocity points is a direction that yields increase of the plasmon frequency as is shown in Fig. \ref{Fig:group_vel}.}

{
Let us study structure of the hyperbolic mode in the upper half-space ($z > 0$). The electric field of the mode (see Eqs. \eqref{Eq:TE}, \eqref{Eq:TM}) takes form
\begin{align}
\mathbf{E}(\mathbf{r}) & = \left[\mathbf{e}'_y \, E_2  - \frac{1}{\omega\varepsilon_0}\left( \mathbf{e}_z k_{\rho} - i \mathbf{e}'_x  \gamma_0 \right) H_2\right] e^{- \gamma_0 z}  e^{i k_{\rho} x  }.
\end{align}
From the boundary conditions (Eq. \eqref{Eq:boundary_cond}) it follows
\begin{align*}
-\frac{2 i \gamma_0}{\omega\mu_0} E_0 = \tilde{\sigma}_{yy} E_0  + \frac{i\gamma_0}{\omega\varepsilon_0}\tilde{\sigma}_{yx}  H_0 \quad \Rightarrow \quad H_0 = - \frac{\omega\varepsilon_0}{i\gamma_0 \tilde{\sigma}_{yx} } \left(\frac{2 i \gamma_0}{\omega\mu_0} +  \tilde{\sigma}_{yy}\right) E_0.
\end{align*}
Thus, the electric field of the mode has following structure
\begin{align*}
\mathbf{E}_2 & = \mathbf{e}'_y  +    \mathbf{e}_z   \left(\frac{2 q}{\omega\mu_0 \tilde{\sigma}_{yx}} +   \frac{q\tilde{\sigma}_{yy}}{i\gamma_0 \tilde{\sigma}_{yx} }\right)  - \mathbf{e}'_x   \left(\frac{2 i \gamma_0}{\omega\mu_0 \tilde{\sigma}_{yx}} +    \frac{\tilde{\sigma}_{yy}}{ \tilde{\sigma}_{yx} }\right).
\end{align*}
We are interested in the mode structure along the hyperbola asymptotes, $\varphi = \phi$ (see Eq. \eqref{Eq:asymptote}), when $\gamma_0 \approx q \gg k_0$. Thus, the mode structure can be further simplified
\begin{align}
&\mathbf{E}_2 \approx   \frac{ \mathbf{e}_z q}{ \tilde{\sigma}_{yx}}  \frac{2 }{\omega\mu_0}    - \frac{ \mathbf{e}'_x}{ \tilde{\sigma}_{yx}}   \frac{2 i \gamma_0}{\omega\mu_0} =  \frac{2 }{\omega\mu_0 \tilde{\sigma}_{yx}} \left( \mathbf{e}_z q - i\mathbf{e}'_x \gamma_0\right) \approx   \frac{2 \gamma_0 }{\omega\mu_0 \tilde{\sigma}_{yx}} \left( \mathbf{e}_z - i\mathbf{e}'_x \right) =  \frac{2 \gamma_0 }{\omega\mu_0 \tilde{\sigma}_{yx}} \left( \mathbf{e}_z - i\mathbf{e}_x \cos \phi - i \mathbf{e}_y \sin\phi \right).
\end{align}
Here we took into account that both $\tilde{\sigma}_{yy}/\tilde{\sigma}_{yx}$ and $\eta_0 \sigma$ are typically of order 1 or less. Assuming that the angle $\phi$ is restricted to the first quadrant, we can define the hyperbolic modes in each of the four quadrants as follows:
\begin{align}
\mathbf{e}_1^m & =- i\mathbf{e}_x \cos \phi - i \mathbf{e}_y \sin\phi +  \mathbf{e}_z, \\
\mathbf{e}_2^m & = i\mathbf{e}_x \cos \phi - i \mathbf{e}_y \sin\phi +  \mathbf{e}_z, \\
\mathbf{e}_3^m & = i\mathbf{e}_x \cos \phi + i \mathbf{e}_y \sin\phi +  \mathbf{e}_z, \\
\mathbf{e}_4^m & = -i\mathbf{e}_x \cos \phi + i \mathbf{e}_y \sin\phi +  \mathbf{e}_z.
\end{align}}

\begin{figure}[h!]
	\centering
	\includegraphics[width=\linewidth]{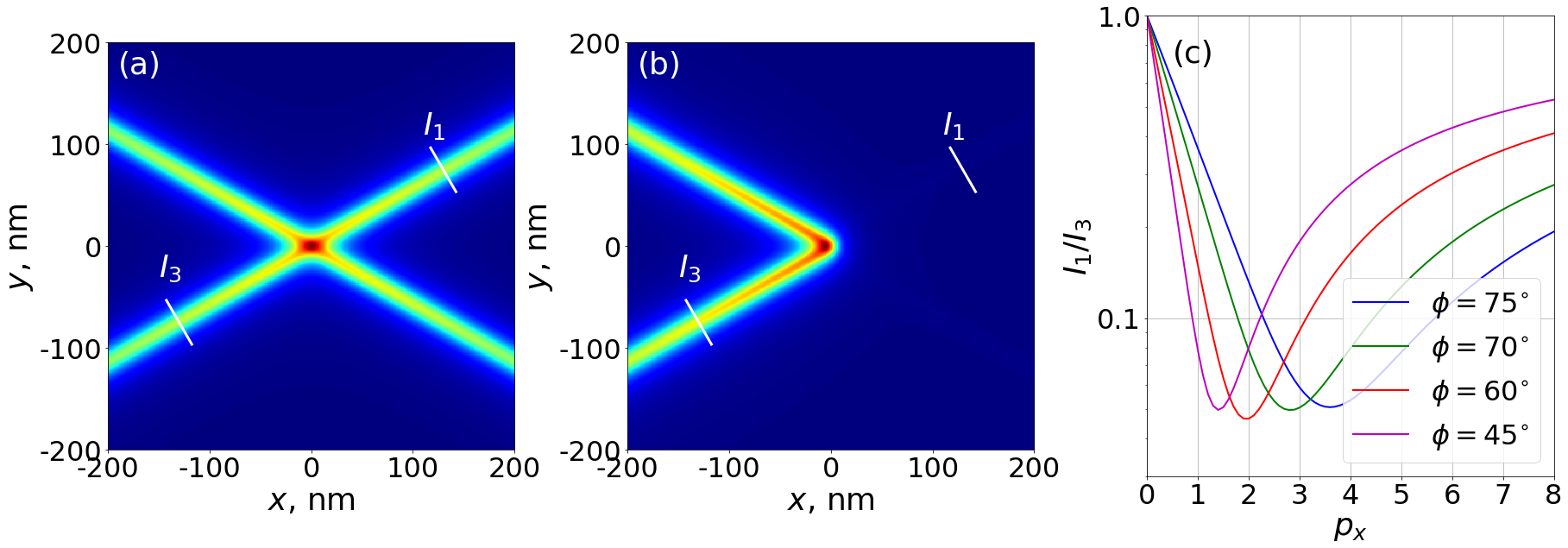}
	\caption{ { (a,b) Electric field, $|\mathbf{E}| = |-\nabla \Phi|$, of the electrostatic plasmon induced in a hyperbolic material ($\sigma''_x =2.85$ mS, $\sigma''_y =- 0.95$ mS, $\phi = 60^{\circ}$) by an electric dipole (a) $\mathbf{p} = - i \mathbf{e}_z$ A$\cdot$m, (b) $\mathbf{p} = 2 \mathbf{e}_x - i \mathbf{e}_z$ A$\cdot$m. $\Phi$ is defined by Eq. \eqref{Eq:Phi_tot_above}. (c) Ratio of intensities, $I_1/I_3$, carried by electrostatic plasmons in the first and third quadrants through detectors (white lines in panels (a,b)).}}
	\label{fig2}
\end{figure}

\begin{figure}[h!]
	\centering
	\includegraphics[width=\linewidth]{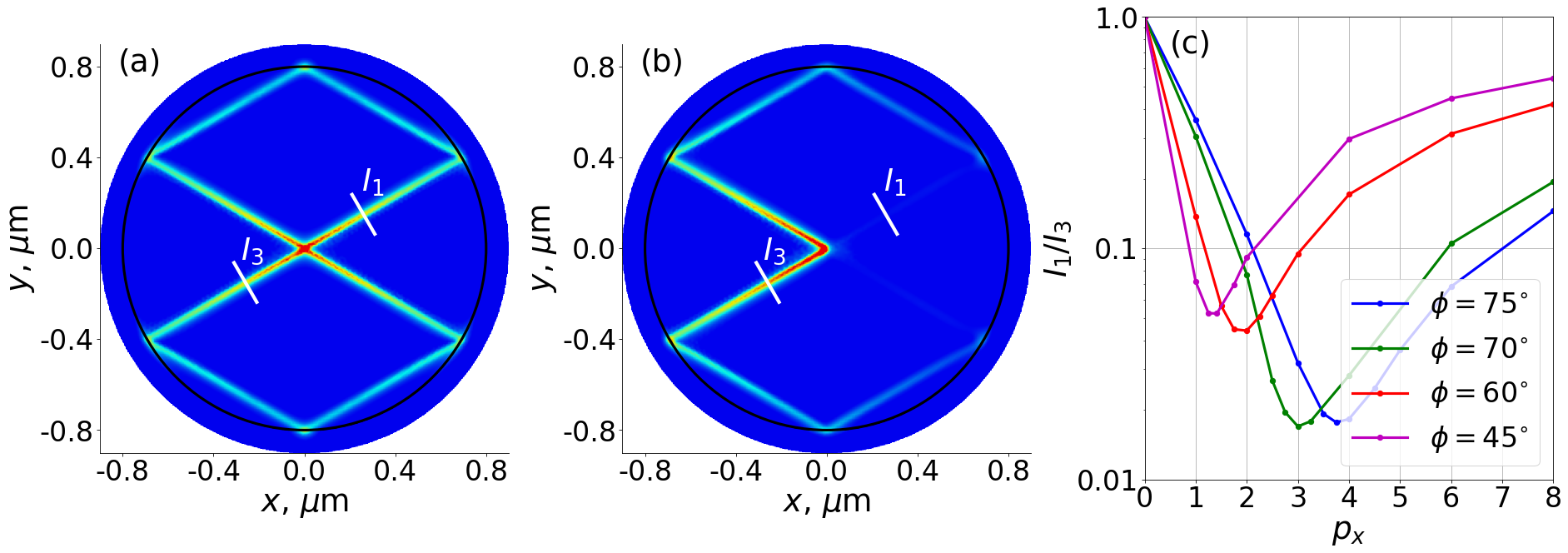}
	\caption{ { (a,b) Electric field of the plasmons induced in a disk of a hyperbolic material ($\sigma''_x =2.85$ mS, $\sigma''_y =- 0.95$ mS, $\phi = 60^{\circ}$) by an electric dipole (a) $\mathbf{p} = - i \mathbf{e}_z$ A$\cdot$m, (b) $\mathbf{p} = 2 \mathbf{e}_x - i \mathbf{e}_z$ A$\cdot$m. (c) Ratio of intensities, $I_1/I_3$, carried by the plasmons in the first and third quadrants through detectors (white lines in panels (a,b)). The calculations were done using COMSOL. The disk radius is 800 nm.}}
	\label{fig2_comsol}
\end{figure}

{As a next step, we calculated an electric field, $\mathbf{E} = - \nabla \Phi$ (where $\Phi$ is defined by Eq. \eqref{Eq:Phi_tot_above}), induced in the hyperbolic material by an electric dipole, $\mathbf{p}$. The calculation results are presented in Fig. \ref{fig2}. For comparison we also repeated the calculations of the electric field using COMSOL. The calculation resuts are presented in Fig. \ref{fig2_comsol}. The comparison of the results obtained using the electrostatic approximation (see Fig. \ref{fig2}) with the results obtained using COMSOL (see Fig. \ref{fig2_comsol}) shows a very good agreement. Thus we can conclude that the electrostatic approximations provides a very good description of the uni-directional excitation of the surface plasmons in the hyperbolic 2D material. }

\section{{Electrostatic potential induced in the hyperbolic 2D material. Analytical approximation of the integral \eqref{Eq:phi_tot_2D} }}

\subsection{{Normalized wavevectors}}

Let us introduce normalized wavevectors
\begin{equation}
\tilde{q}_x^2 = 2 q_x^2 \sin^2 \phi, \quad \tilde{q}_y^2 = 2 q_y^2 \cos^2 \phi.
\end{equation}
In this case,
\begin{align}
t_{\mathbf{q}} = \left(1 - v_c(\mathbf{q}) \chi_0(\mathbf{q})\right)^{-1} = \left(1 + \frac{i  v_c(\mathbf{q})}{\omega } \, \left(q_x^2 \sigma_x + q_y^2 \sigma_y \right) \right)^{-1} = \left(1 +  \frac{i  v_c(\mathbf{q})}{\omega } \, \left(q_x^2 \sigma'_x + q_y^2 \sigma'_y \right) - \frac{ v_c(\mathbf{q})}{\omega } \, \left(q_x^2 |\sigma''_x| - q_y^2 |\sigma''_y| \right) \right)^{-1}. \notag
\end{align}
We should take into account that
\begin{align*}
\frac{\sin^2 \phi}{\cos^2\phi} = \left|\frac{\sigma''_x}{\sigma''_y}\right| \quad \Rightarrow \quad \frac{\sin^2 \phi}{|\sigma''_x|} = \frac{\cos^2\phi}{|\sigma''_y|}
\end{align*}
and
\begin{align*}
\frac{1}{\cos^2\phi} = \left|\frac{\sigma''_x}{\sigma''_y}\right| + 1 \qquad \Rightarrow \qquad \frac{\left|\sigma''_y\right| }{\cos^2\phi} = \left|\sigma''_x\right| +\left|\sigma''_y\right| = \frac{\left|\sigma''_x\right| }{\sin^2\phi}.
\end{align*}
Then
\begin{align*}
&q_x^2 |\sigma''_x| - q_y^2 |\sigma''_y| = \frac{\tilde{q}_x^2 |\sigma''_x|}{2 \sin^2 \phi} - \frac{\tilde{q}_y^2 |\sigma''_y|}{2 \cos^2 \phi} = \frac{\left|\sigma''_x\right| +\left|\sigma''_y\right| }{2} \left(\tilde{q}_x^2  - \tilde{q}_y^2 \right)
\end{align*}
and
\begin{align}
t_{\mathbf{q}} & =  \left(1 +  \frac{i }{2 \epsilon_0 \omega |\mathbf{q}| } \, \left(q_x^2 \sigma'_x + q_y^2 \sigma'_y \right) - \frac{ \bar{\sigma}}{2 \epsilon_0 \omega |\mathbf{q}| } \, \left(\tilde{q}_x^2 - \tilde{q}_y^2  \right) \right)^{-1},
\end{align}
where $ \bar{\sigma} = (\left|\sigma''_x\right| +\left|\sigma''_y\right|)/2$.
Let us define new variables
\begin{equation}
q_{\pm} = \tilde{q}_x \pm \tilde{q}_y \qquad \Rightarrow \qquad \tilde{q}_x = \frac{q_+ + q_-}{2}, \qquad \tilde{q}_y = \frac{q_+ - q_-}{2}.
\end{equation}
Using these new variables, we obtain
\begin{align}
\tilde{q}_x^2 - \tilde{q}_y^2 = \frac{q_+^2 + 2 q_+ q_- + q_-^2}{4} - \frac{q_+^2 - 2 q_+ q_- + q_-^2}{4} = q_+ q_-.
\end{align}
Thus,
\begin{align}
t_{\mathbf{q}} & =  \left(1 +  \frac{i }{2 \epsilon_0 \omega |\mathbf{q}| } \, \left(q_x^2 \sigma'_x + q_y^2 \sigma'_y \right) - \frac{ q_+ q_-}{Q_0 |\mathbf{q}| } \,  \right)^{-1},
\end{align}
where $Q_0 = 2\epsilon_0\omega/\bar{\sigma}$,
\begin{align*}
&|\mathbf{q}| = \sqrt{\frac{\tilde{q}_x^2}{2 \sin^2 \phi} + \frac{\tilde{q}_y^2}{2 \cos^2 \phi}} = \sqrt{\frac{(q_+ + q_-)^2 \cos^2 \phi  + (q_+ - q_-)^2 \sin^2\phi}{8 \sin^2 \phi \cos^2 \phi}}  = \sqrt{\frac{q_+^2 + q_-^2 + 2 q_+ q_- \cos 2\phi }{2 \sin^2 2\phi }},
\end{align*}
and
\begin{align*}
q_x^2 \sigma'_x + q_y^2 \sigma'_y &= \frac{\tilde{q}_x^2 \sigma'_x}{2 \sin^2 \phi} + \frac{\tilde{q}_y^2 \sigma'_y}{2\cos^2\phi} = \frac{1}{8} \left( \frac{\sigma'_x}{\sin^2 \phi} (q_+^2 + 2 q_+ q_- + q_-^2) + \frac{\sigma'_y}{\cos^2\phi} (q_+^2 - 2 q_+ q_- + q_-^2)\right) \\
& = \frac{1}{8} \left( \left(\frac{\sigma'_x}{\sin^2 \phi} + \frac{\sigma'_y}{\cos^2\phi}\right) \left( q_+^2 +  q_-^2  \right) + \left(\frac{\sigma'_x}{\sin^2 \phi} - \frac{\sigma'_y}{\cos^2\phi}\right) 2 q_+ q_-\right).
\end{align*}

\subsection{Asymptotic approximation, $q_+ \gg q_-$}
Let us assume that the dominant contribution to the integral comes from the plasmons with wave vectors on hyperbola asymptotes, i.e., $q_x = q \cos\phi$ and $q_y = q \sin\phi$. Then
\begin{equation}
q_- = \tilde{q}_x - \tilde{q}_y = 2 q_x^2 \sin^2 \phi - 2 q_y^2 \cos^2 \phi = 2 q^2 \sin^2 \phi  \cos^2 \phi -  2 q^2 \sin^2 \phi  \cos^2 \phi = 0.
\end{equation}
Thus around asymptotes  $q_+ \gg q_-$ we can simplify even further
\begin{align*}
&|\mathbf{q}| =  \frac{q_+ }{\sqrt{2}|\sin 2\phi|}, \\
& q_x^2 \sigma'_x + q_y^2 \sigma'_y  =  \frac{1}{8}  \left(\frac{\sigma'_x}{\sin^2 \phi} + \frac{\sigma'_y}{\cos^2\phi}\right)  q_+^2.
\end{align*}
Then
\begin{align}
t_{\mathbf{q}} & =  \left(1 +  \frac{i}{8}  \left(\frac{\sigma'_x}{\sin^2 \phi} + \frac{\sigma'_y}{\cos^2\phi}\right)  q_+  \frac{ \sqrt{2} |\sin 2\phi| }{2 \epsilon_0 \omega } \, - \frac{ \sqrt{2} q_- |\sin 2\phi|}{Q_0 } \,  \right)^{-1} \notag \\
& = \left(1 + \frac{i}{8}  \left(\frac{\sigma'_x}{\bar{\sigma}\sin^2 \phi} + \frac{\sigma'_y}{\bar{\sigma}\cos^2\phi}\right)   \frac{  q_+  }{q_0} \,  - \frac{  q_- }{q_0 } \,  \right)^{-1} \notag \\
&  = - q_0 \left(q_- - q_0 - i\gamma_0 q_+\right)^{-1},
\end{align}
where 
\begin{equation}
\begin{split}
q_0 &= Q_0/\sqrt{2} |\sin 2\phi|,\\
\gamma_0 &= \frac{1}{8}  \left(\frac{\sigma'_x}{\bar{\sigma}\sin^2 \phi} + \frac{\sigma'_y}{\bar{\sigma}\cos^2\phi}\right).
\end{split}
\end{equation}

\subsection{Electrostatic potential. Numerical integration }
{The electrostatic potential in the point $(x, y)$ at the surface of the hyperbolic material, due to a general electric dipole $\mathbf{p} = p_x \mathbf{e}_x + p_y \mathbf{e}_y + p_z \mathbf{e}_z$, can be calculated as}
\begin{align} \label{Eq:Phi}
\Phi = p_x \Phi_x + p_y \Phi_y + p_z \Phi_z.
\end{align}

\begin{align}
\Phi_x & = - i \iint \frac{dq_x dq_y}{(2\pi)^2} \,v_c(\mathbf{q}) t_{\mathbf{q}}  e^{i (q_x x + q_y y)} e^{ - |\mathbf{q}| |z_0|}  q_x  \notag \\
& = -\frac{i}{8\pi^2\epsilon_0} \int_{0}^{2\pi} d\theta \cos\theta \int_{0}^{\infty} dq \, q\, t_{q,\theta} \, e^{i q \left(x\cos\theta + y\sin\theta\right)} e^{ - q |z_0|} ,
\end{align}
\begin{align}
\Phi_y & = - i \iint \frac{dq_x dq_y}{(2\pi)^2} \,v_c(\mathbf{q}) t_{\mathbf{q}}  e^{i (q_x x + q_y y)} e^{ - |\mathbf{q}| |z_0|}  q_y  \notag \\
& = -\frac{i}{8\pi^2\epsilon_0} \int_{0}^{2\pi} d\theta \sin\theta \int_{0}^{\infty} dq \, q\, t_{q,\theta} \, e^{i q \left(x\cos\theta + y\sin\theta\right)} e^{ - q |z_0|} ,
\end{align}
\begin{align}
\Phi_z & = -\iint \frac{dq_x dq_y}{(2\pi)^2} \,v_c(\mathbf{q}) t_{\mathbf{q}}  e^{i (q_x x + q_y y)} e^{ - |\mathbf{q}| |z_0|}  |\mathbf{q}| \mathrm{sign}(z_0) \\
& = -\frac{\mathrm{sign}(z_0)}{8\pi^2\epsilon_0} \int_{0}^{2\pi} d\theta \int_{0}^{\infty} dq \, q\, t_{q,\theta} \, e^{i q \left(x\cos\theta + y\sin\theta\right)} e^{ - q |z_0|} ,
\end{align}
where
\begin{align}
t_{\mathbf{q}} =  \left(1 + \frac{i }{2 \epsilon_0 q \omega } \, \left(q_x^2 \sigma_x + q_y^2 \sigma_y \right) \right)^{-1} =  \left(1 + \frac{i q }{2 \epsilon_0 \omega } \, \left( \sigma_x \cos^2 \theta  +\sigma_y \sin^2\theta\right) \right)^{-1} .
\end{align}

\subsection{Electrostatic potential due to the $z$-polarized dipole, i.e., $\mathbf{p} = \mathbf{e}_z$. Analytic approximation. Integration over the first quadrant in $\mathbf{q}$ space.  }
\label{Sec:approx_Phiz}

We start with integral
\begin{align}
\Phi_z & = -\iint \frac{dq_x dq_y}{(2\pi)^2} \,v_c(\mathbf{q}) t_{\mathbf{q}}  e^{i (q_x x + q_y y)} e^{ - |\mathbf{q}| |z_0|}  |\mathbf{q}| \mathrm{sign}(z_0).
\end{align}
Let us consider integration over the first quadrant first
\begin{align}
\Phi_{z, 1st} & = -\int_{0}^{\infty} dq_x\int_{0}^{\infty} \, dq_y \frac{ 1}{(2\pi)^2} \,v_c(\mathbf{q}) t_{\mathbf{q}}  e^{i (q_x x + q_y y)} e^{ - |\mathbf{q}| |z_0|}  |\mathbf{q}| \mathrm{sign}(z_0) \notag \\
& = -\frac{ 1}{(2\pi)^2 2\sin\phi \cos\phi} \int_{0}^{\infty} d\tilde{q}_x\int_{0}^{\infty} \, d\tilde{q}_y  \frac{1}{2\epsilon_0}\,t_{\tilde{\mathbf{q}}}  e^{i (\tilde{q}_x \tilde{x} + \tilde{q}_y \tilde{y})} e^{ - |\mathbf{q}| |z_0|}   \mathrm{sign}(z_0),   \label{Eq:PhiZ_exact_1st}
\end{align}
where $\tilde{x} = x/\sqrt{2}\sin{\phi}$, $\tilde{y} = y/\sqrt{2}\cos\phi$.
In order to change integration variables to $q_+$, $q_-,$ we need to take into account that
\begin{equation}
\left|\frac{\partial\left(\tilde{q}_x, \tilde{q}_y\right)}{\partial\left(q_+, q_-\right)}\right| = \left|
\begin{array}{cc}
\partial \tilde{q}_x/ \partial q_+ & \partial \tilde{q}_x/ \partial q_- \\
\partial \tilde{q}_y/ \partial q_+ & \partial \tilde{q}_y/ \partial q_-
\end{array}
\right| = \left|
\begin{array}{cc}
1/2 & 1/2 \\
1/2 & -1/2
\end{array}
\right|= \frac{1}{2}.
\end{equation}
In order to define integration limits, we take into account that the integration domain is defined by the boundaries $\tilde{q}_x = 0$, $\tilde{q}_y = 0$, which imposes the following restriction on the boundaries of the new domain: $q_- = q_+$, $q_- = - q_+$. Moreover, $q_+ \geq 0$. Thus we obtain
\begin{align}
\Phi_{z,1st} & = \frac{ q_0 \mathrm{sign}(z_0)}{16\epsilon_0\pi^2 \sin 2\phi } \int_{0}^{\infty} dq_+\int_{-q_+}^{q_+} \, dq_-  \, \frac{e^{i ( (q_+ + q_-) \tilde{x}/2 + (q_+ - q_-) \tilde{y}/2)}}{q_- - q_0 - i\gamma_0 q_+} e^{ - q_+ |z_0|/(\sqrt{2}|\sin 2\phi|)}  \notag \\
&= \frac{ q_0 \mathrm{sign}(z_0)}{16\epsilon_0\pi^2 \sin 2\phi } \int_{0}^{\infty} dq_+ e^{i q_+ r_+ }e^{ - q_+ |\tilde{z}_0|}\int_{-q_+}^{q_+} \, dq_-  \, \frac{e^{i  q_-r_-}}{q_- - q_0 - i\gamma_0 q_+} ,
\end{align}
where $r_{\pm} = (\tilde{x}\pm\tilde{y})/2$, and $\tilde{z}_0 = z_0/(\sqrt{2}|\sin 2\phi|)$.

Let us first consider inner integral
\begin{equation}
I_+ = \int_{-q_+}^{q_+} \, dq_-  \, \frac{e^{i  q_-r_-}}{q_- - q_0 - i\gamma_0 q_+}.
\end{equation}
Note that the integrand in $I_+$ has a pole when $q_- = q_0$. However, the pole is within the integration limits only when
\begin{equation} \label{Eq:qplus}
q_+ \geq q_0.
\end{equation}
Assuming that \eqref{Eq:qplus} is true, we obtain
\begin{align}
I_+ \approx \int_{-\infty}^{\infty} \, dq_-  \, \frac{e^{i  q_-r_-}}{q_- - q_0 - i\gamma_0 q_+} +  \int_{C_{\infty}} \, dq_-  \, \frac{e^{i  q_-r_-}}{q_- - q_0 - i\gamma_0 q_+} = 2\pi i \theta(r_-) e^{i (q_0 + i\gamma_0 q+)r_-},
\end{align}
where the Heaviside function, $\theta$, ensures that the integrand is zero along the infinity contour, $C_{\infty}$.

The outer integral then turns into
\begin{align} \label{Eq:Phi1st_approx}
\Phi_{z, appr, 1st} & = \theta(r_-)  \mathrm{sign}(z_0)\frac{i q_0 e^{i q_0 r_-}}{8\epsilon_0\pi \sin 2\phi} \int_{q_0}^{\infty} dq_+ e^{i q_+ r_+ }e^{ - q_+ \left(|\tilde{z}_0| + \gamma_0 |r_-|\right)} \notag \\
& = -\theta(r_-)  \mathrm{sign}(z_0)\frac{i q_0 e^{i q_0 r_-}}{8\epsilon_0\pi \sin 2\phi } \frac{ e^{i q_0 r_+ }e^{ - q_0 \left(|\tilde{z}_0| + \gamma_0 |r_-|\right)}  }{i r_+ - \left(|\tilde{z}_0| + \gamma_0 |r_-|\right)} \notag \\
& = -\theta(\tilde{x}-\tilde{y})  \mathrm{sign}(z_0)\frac{q_0 e^{i q_0 (\tilde{x}-\tilde{y})/2}}{4\epsilon_0\pi \sin 2\phi} \frac{ e^{i q_0 (\tilde{x}+\tilde{y})/2 }e^{ - q_0 \left(|\tilde{z}_0| + \gamma_0 |\tilde{x}-\tilde{y}|/2\right)}  }{\tilde{x}+\tilde{y} + i \left(2|\tilde{z}_0| + \gamma_0 |\tilde{x}-\tilde{y}|\right)} .
\end{align}

Let us take into account that
\begin{equation}
\tilde{x} + \tilde{y} = \frac{x}{\sqrt{2} \sin\phi} + \frac{y}{\sqrt{2} \cos\phi} = \frac{1}{\sqrt{2} \sin\phi} \left(x + y\tan\phi\right),
\end{equation}
and that hyperbolic plasmons carry energy along the direction of the group velocity, which in the first quadrant points along the direction
\begin{equation} \label{Eq:hyperbolic_energy}
\frac{y}{x} = - \sqrt{\left|\frac{\sigma''_y}{\sigma''_x}\right|} = - \frac{1}{\tan\phi}.
\end{equation}
Thus $\tilde{x} + \tilde{y} \approx 0$ and $e^{i q_0 (\tilde{x}+\tilde{y})/2} \approx 1$ along the direction of the hyperbolic beam propagation, and we obtain
\begin{align}
\Phi_{z, appr, 1st} & \approx -\theta(\tilde{x}-\tilde{y})  \mathrm{sign}(z_0)\frac{q_0 e^{i q_0 (\tilde{x}-\tilde{y})/2}}{4\epsilon_0\pi \sin 2\phi} \frac{ e^{ - q_0 \left(|\tilde{z}_0| + \gamma_0 |\tilde{x}-\tilde{y}|/2\right)}  }{\tilde{x}+\tilde{y} + i \left(2|\tilde{z}_0| + \gamma_0 |\tilde{x}-\tilde{y}|\right)}.
\end{align}

As a next step, let us assume that we study plasmons at the frequency $\nu = 46$ THz, and the conductivities are $\sigma_x = 0.05 + i 2.85$ mS and $\sigma_y = 0.015 -i 0.95$ mS, which corresponds to $\phi = \pi/3$. This corresponds to
\begin{equation}
Q_0 = \frac{4 \epsilon_0 \omega}{\left|\sigma''_x\right| +\left|\sigma''_y\right|} = 2.6 \mu\mathrm{m}^{-1}, \qquad q_0 = \frac{Q_0}{\sqrt{2}|\sin 2\phi|} = 2.16 \mu\mathrm{m}^{-1}.
\end{equation}
Thus
\begin{equation}
q_0 |\tilde{z}_0| = \frac{q_0}{\sqrt{2}|\sin 2\phi|} \, |z_0| = 1.78\times 10^{6} |z_0| \ll 1 \qquad \mathrm{if~}|z_0| < 10 \mathrm{~nm}.
\end{equation}
This means that $e^{ - q_0 |\tilde{z}_0|} \approx 1$. Moreover, the relaxation parameter $\gamma_0 = 0.008,$ and thus
\begin{equation}
q_0 \gamma_0 |r_-| = 1.5\times 10^4 |r_-| \ll 1 \qquad \mathrm{if~}|r_-| < 10~\mu\mathrm{m}.
\end{equation}
This means that $e^{ - q_0 \gamma_0 |\tilde{x}-\tilde{y}|/2} \approx 1$. Thus,
\begin{align}
\Phi_{z, appr, 1st} \equiv \Phi_1 & \approx \frac{ i  q_0 \, \mathrm{sign}(z_0) \,\theta(\tilde{x}-\tilde{y}) }{4\epsilon_0\pi \sin 2\phi } \frac{ e^{i q_0 (\tilde{x}-\tilde{y})/2}}{2|\tilde{z}_0| + \gamma_0 |\tilde{x}-\tilde{y}|}
\end{align}

\begin{figure}[h!]
	\centering
\includegraphics[width=0.4\textwidth]{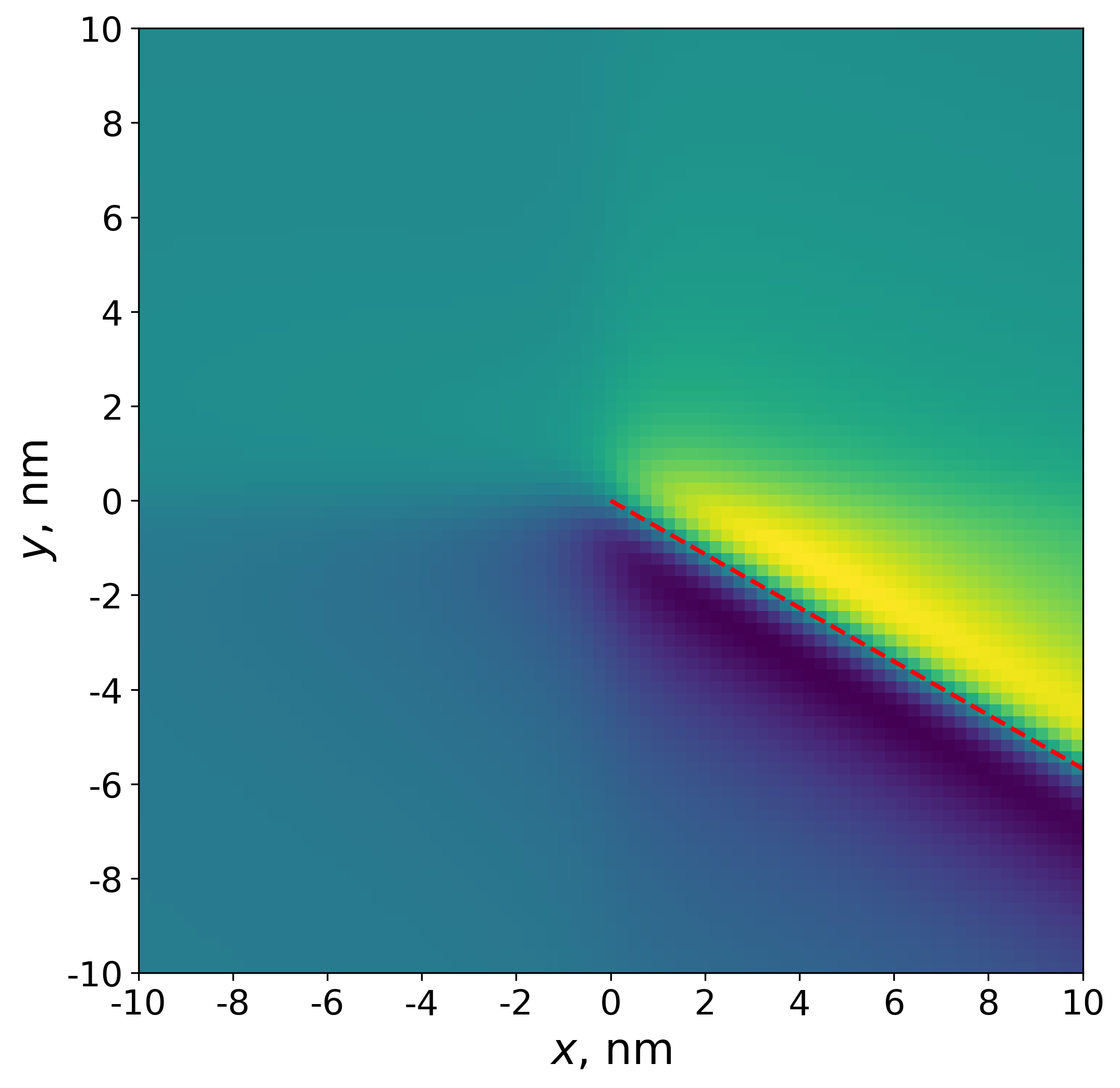}
\includegraphics[width=0.4\textwidth]{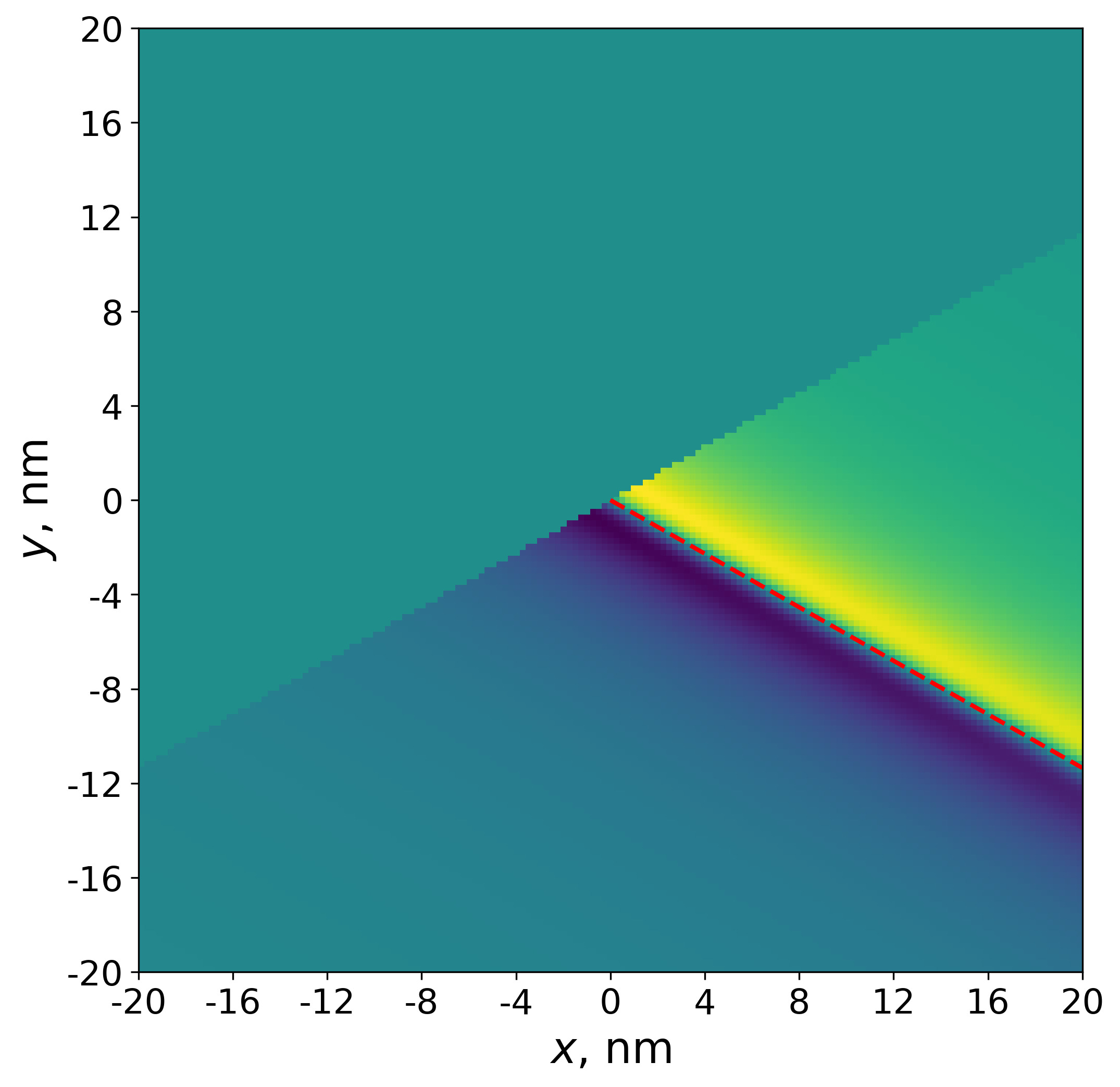}
	\caption{Real part of electrostatic potential, $\Phi_{1st}$, calculated using \textbf{(a)} the exact Eq. \eqref{Eq:PhiZ_exact_1st}, and \textbf{(b)} the approximate Eq. \eqref{Eq:Phi1st_approx}. Dashed red line shows the direction of hyperbolic plasmon energy flow, calculated using Eq. \eqref{Eq:hyperbolic_energy}. $\nu = 46$ THz,  $\sigma''_x =2.85$ mS and $\sigma''_y =- 0.95$ mS. }
	\label{Fig:RePhiZ_comparison_1st}
\end{figure}

\begin{figure}[h!]
	\centering
\includegraphics[width=0.4\textwidth]{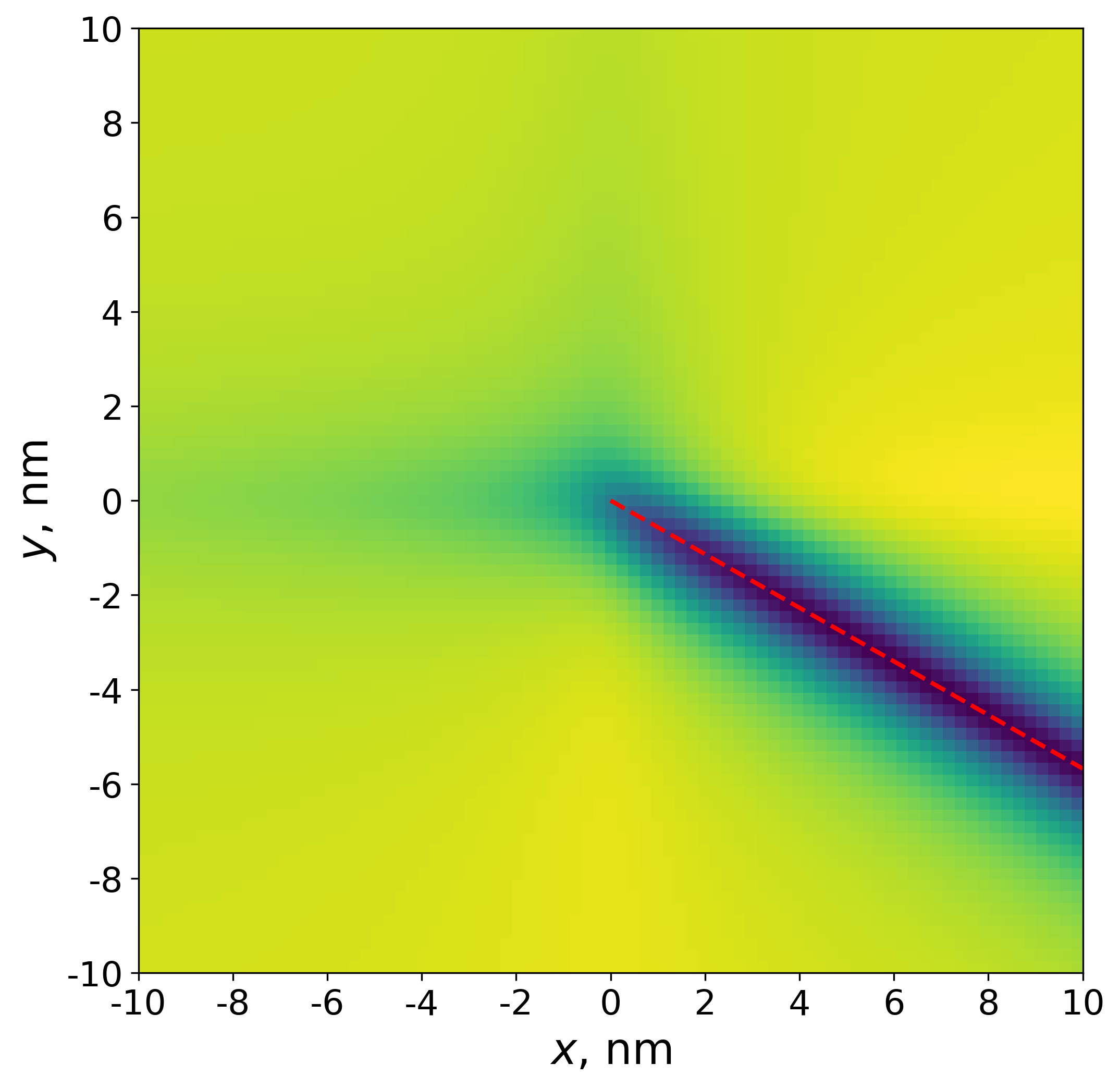}
\includegraphics[width=0.4\textwidth]{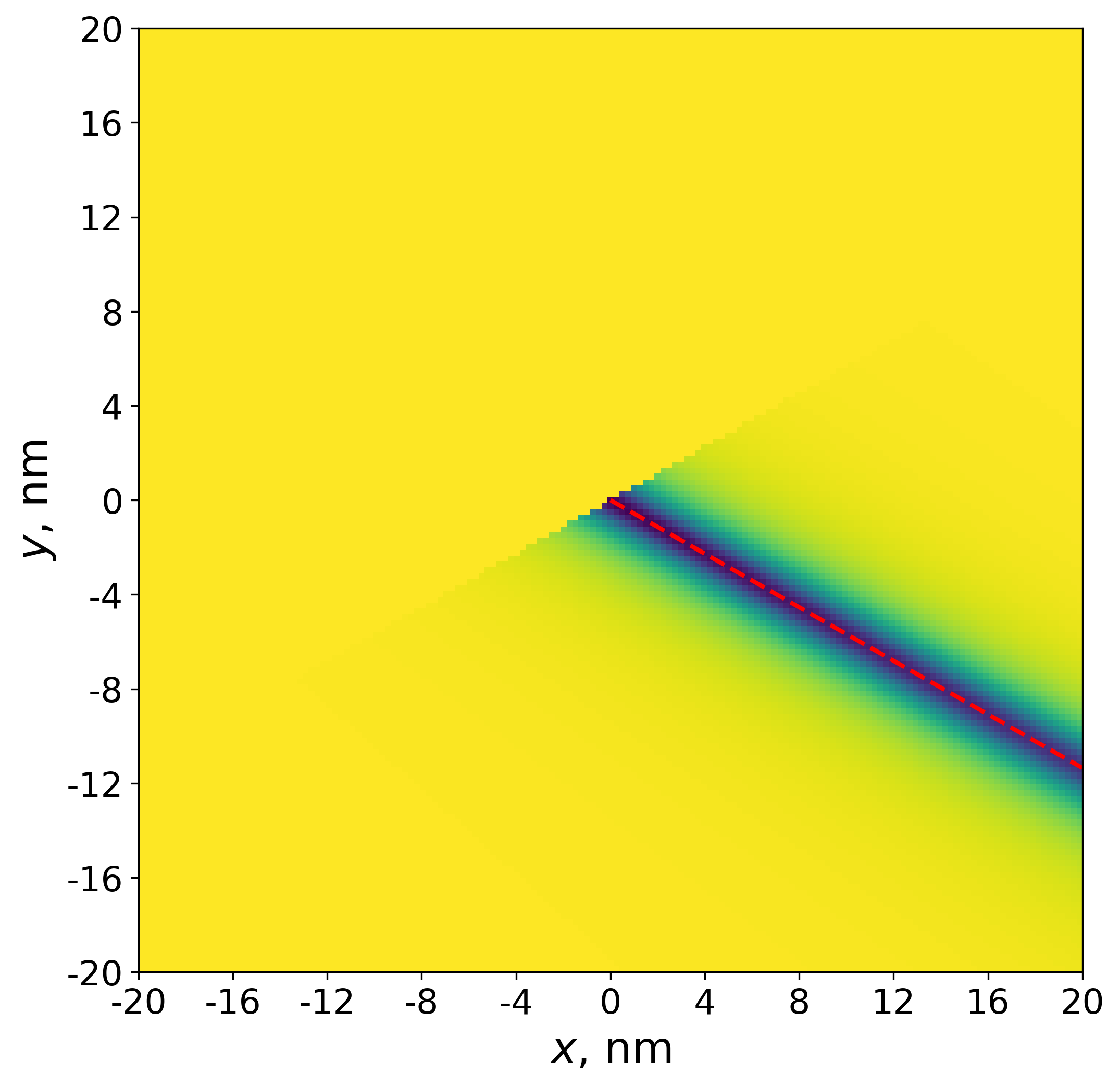}
	\caption{Imaginary part of electrostatic potential, $\Phi_{1st}$, calculated using \textbf{(a)} the exact Eq. \eqref{Eq:PhiZ_exact_1st}, and \textbf{(b)} the approximate Eq. \eqref{Eq:Phi1st_approx}. Dashed red line shows the direction of hyperbolic plasmon energy flow, calculated using Eq. \eqref{Eq:hyperbolic_energy}. $\nu = 46$ THz,  $\sigma''_x =2.85$ mS and $\sigma''_y =- 0.95$ mS. }
	\label{Fig:ImPhiZ_comparison_1st}
\end{figure}

\begin{figure}[h!]
	\centering
\includegraphics[width=0.4\textwidth]{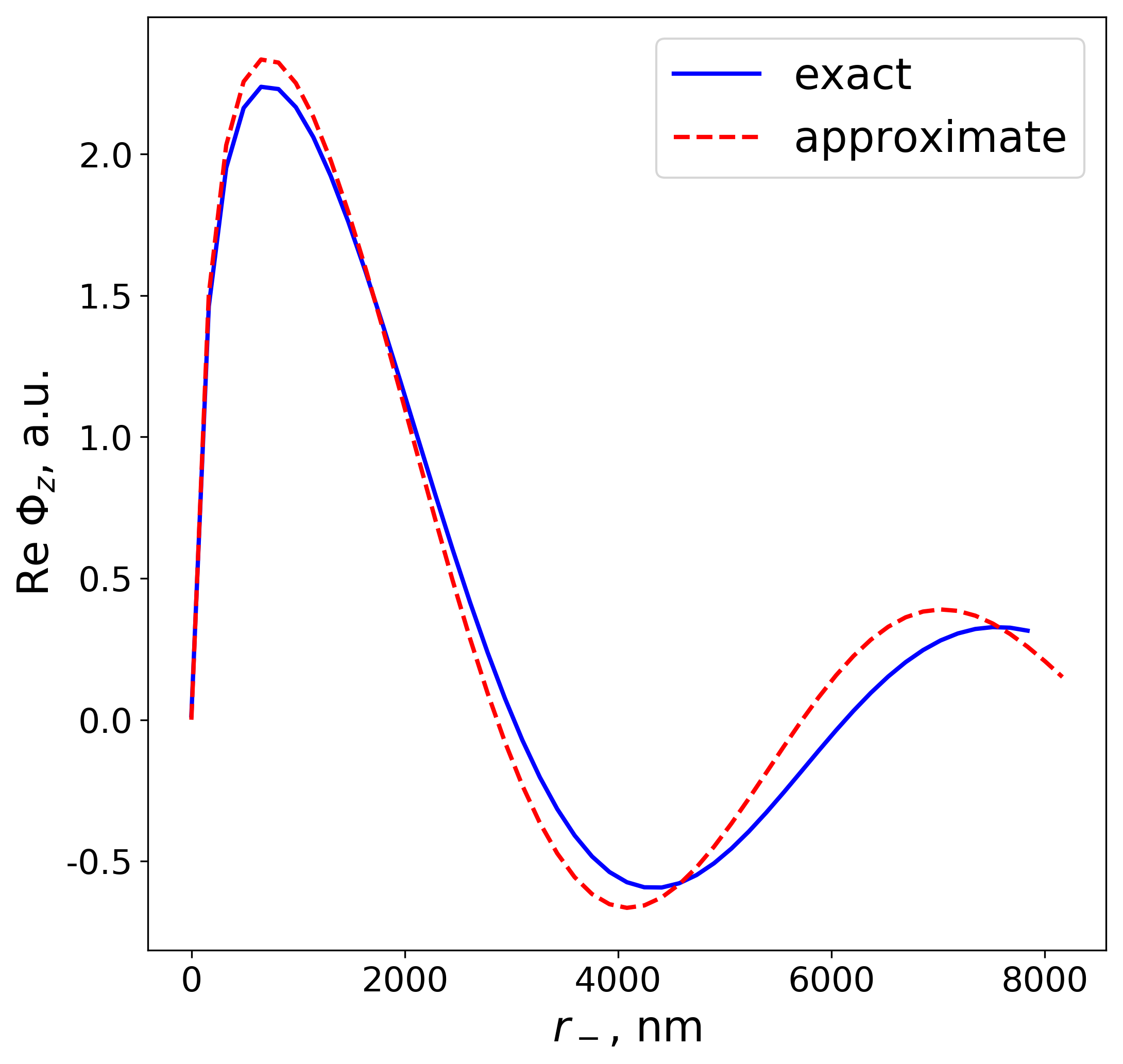}
\includegraphics[width=0.4\textwidth]{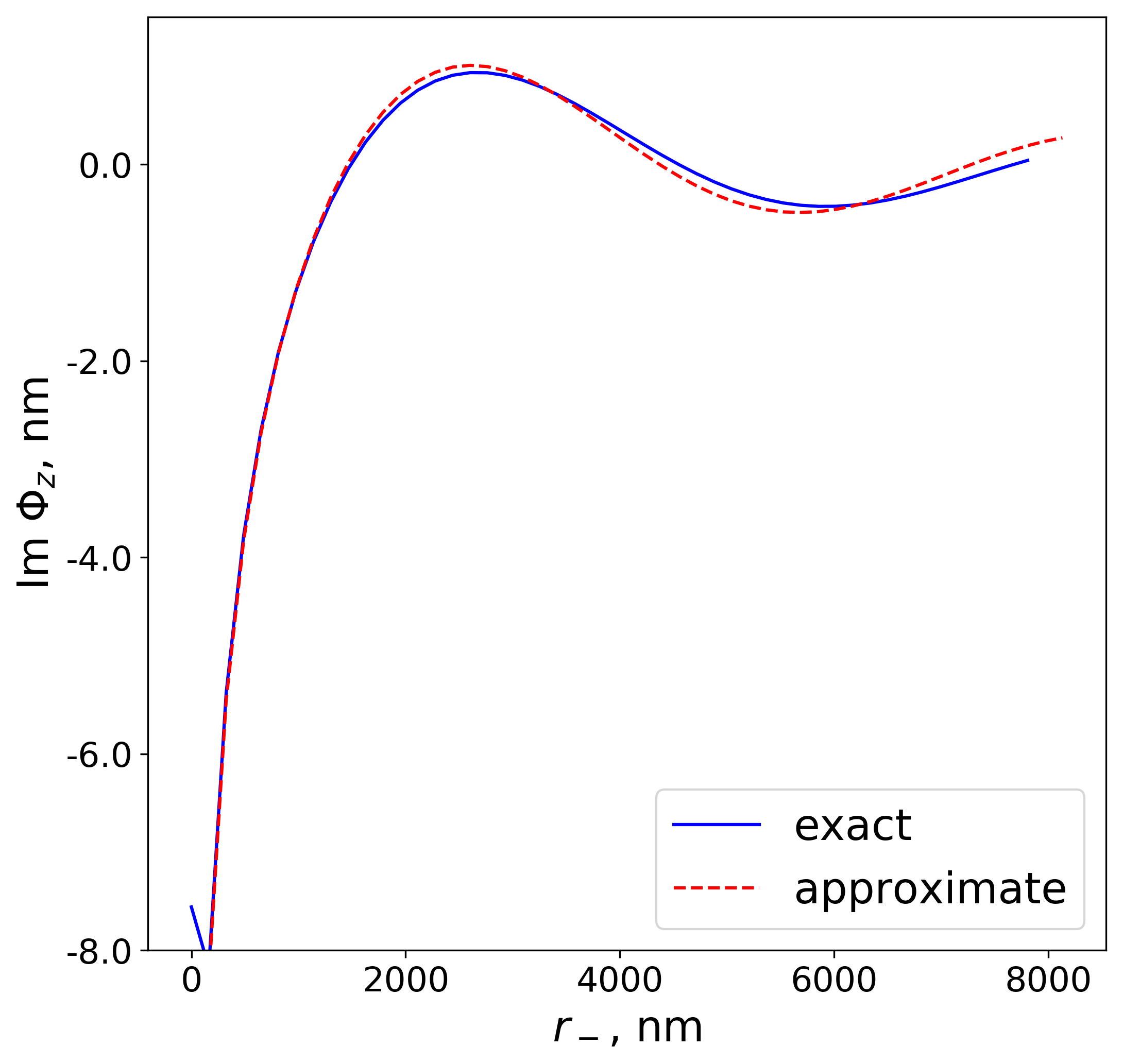}
	\caption{(a) Real and (b) imaginary parts of electrostatic potential, $\Phi_{1st}$, calculated using the exact Eq. \eqref{Eq:PhiZ_exact_1st}, and the approximate Eq. \eqref{Eq:Phi1st_approx}. $\nu = 46$ THz,  $\sigma''_x =2.85$ mS and $\sigma''_y =- 0.95$ mS. }
	\label{Fig:ReImPhiZ_line}
\end{figure}

Comparison between the approximate Eq. \eqref{Eq:Phi1st_approx} and exact Eq. \eqref{Eq:PhiZ_exact_1st} is presented in Figs. \ref{Fig:RePhiZ_comparison_1st}, \ref{Fig:ImPhiZ_comparison_1st}. As one can see, integration of the first quadrant in $\mathbf{q}$ space produces only one ray in the forth quadrant in the coordinate space.

\subsection{Electrostatic potential due to the $x$-polarized and $y$-polarized dipoles, i.e., $\mathbf{p} = \mathbf{e}_x$ or $\mathbf{p} = \mathbf{e}_y$. Analytic approximation. Integration over the first quadrant in $\mathbf{q}$ space.  }
We start with the integral
\begin{align}
\Phi_{x,y} & = - i\iint \frac{dq_x dq_y}{(2\pi)^2} \,v_c(\mathbf{q}) t_{\mathbf{q}}  e^{i (q_x x + q_y y)} e^{ - |\mathbf{q}| |z_0|}  q_{x,y}.
\end{align}
Then we can define integration over the first quadrant as
\begin{align}
\Phi_{x, appr, 1st} & =  - i \int_{0}^{\infty} dq_x\int_{0}^{\infty} \, dq_y \frac{v_c(\mathbf{q}) t_{\mathbf{q}}}{(2\pi)^2}   e^{i (q_x x + q_y y)} e^{ - |\mathbf{q}| |z_0|}  |\mathbf{q}| \cos\phi \notag = \frac{ i \cos\phi \, \Phi_{z, 1st, approx}}{\mathrm{sign}(z_0)}, \\
\Phi_{y, appr, 1st} & =  - i \int_{0}^{\infty} dq_x\int_{0}^{\infty} \, dq_y \frac{v_c(\mathbf{q}) t_{\mathbf{q}}}{(2\pi)^2}   e^{i (q_x x + q_y y)} e^{ - |\mathbf{q}| |z_0|}  |\mathbf{q}| \sin\phi \notag = \frac{i \sin\phi \Phi_{z, 1st, approx} }{\mathrm{sign}(z_0)}
\end{align}
Thus, the approximated electrostatic potential due to a general electric dipole, $\mathbf{p} = p_x \mathbf{e}_x + p_y \mathbf{e}_y + p_z \mathbf{e}_z$ is equal to
\begin{equation}
\Phi_{appr, 1st}(x,y) = \mathrm{sign}(z_0) \Phi_1(x, y) \left(i p_x \cos\phi + i p_y \sin\phi + \mathrm{sign}(z_0)  p_z\right)
\end{equation}

\subsection{Electrostatic potential. Second quadrant in $\mathbf{q}$ space.  }
\begin{align}
\Phi_{z, 2nd} (x, y) & = -\int_{-\infty}^{0} dq_x\int_{0}^{\infty} \, dq_y \frac{ 1}{(2\pi)^2} \,v_c(\mathbf{q}) t_{\mathbf{q}}  e^{i (q_x x + q_y y)} e^{ - |\mathbf{q}| |z_0|}  |\mathbf{q}| \mathrm{sign}(z_0) \notag \\ & =  -\int_{0}^{\infty} dq_x\int_{0}^{\infty} \, dq_y \frac{ 1}{(2\pi)^2} \,v_c(\mathbf{q}) t_{\mathbf{q}}  e^{i (- q_x x + q_y y)} e^{ - |\mathbf{q}| |z_0|}  |\mathbf{q}| \mathrm{sign}(z_0) = \Phi_{z, 1st} (-x, y).
\end{align}

In order to calculate $\Phi_{x, 2nd}$ and $\Phi_{y, 2nd}$, we should take into account that in the second quadrant $q_x = - q \cos\phi$, $q_y = q \sin\phi$. Then
\begin{align}
\Phi_{x, appr, 2nd} & =   -\frac{ i \cos\phi \, \Phi_{z, 2nd, approx}}{\mathrm{sign}(z_0)}, \qquad \Phi_{y, appr, 2nd}  = \frac{i \sin\phi \Phi_{z, 2nd, approx} }{\mathrm{sign}(z_0)},
\end{align}
and
\begin{equation}
\Phi_{appr, 2nd}(x,y) = \mathrm{sign}(z_0) \Phi_1(-x, y) \left(- i p_x \cos\phi + i p_y \sin\phi + \mathrm{sign}(z_0)  p_z\right)
\end{equation}

\subsection{Electrostatic potential. Third quadrant in $\mathbf{q}$ space.  }
\begin{align}
\Phi_{z, 3rd} (x, y) & = -\int_{-\infty}^{0} dq_x\int_{-\infty}^{0} \, dq_y \frac{ 1}{(2\pi)^2} \,v_c(\mathbf{q}) t_{\mathbf{q}}  e^{i (q_x x + q_y y)} e^{ - |\mathbf{q}| |z_0|}  |\mathbf{q}| \mathrm{sign}(z_0) \notag \\ & =  -\int_{0}^{\infty} dq_x\int_{0}^{\infty} \, dq_y \frac{ 1}{(2\pi)^2} \,v_c(\mathbf{q}) t_{\mathbf{q}}  e^{i (- q_x x - q_y y)} e^{ - |\mathbf{q}| |z_0|}  |\mathbf{q}| \mathrm{sign}(z_0) = \Phi_{z, 1st} (-x, -y).
\end{align}

In order to calculate $\Phi_{x, 3rd}$ and $\Phi_{y, 3rd}$, we should take into account that in the third quadrant $q_x = - q \cos\phi$, $q_y = - q \sin\phi$.
\begin{align}
\Phi_{x, appr, 3rd} & =   -\frac{ i \cos\phi\, \Phi_{z, 3rd, approx}}{\mathrm{sign}(z_0)}, \qquad \Phi_{y, appr, 3rd}  = -\frac{i \sin\phi \Phi_{z, 3rd, approx} }{\mathrm{sign}(z_0)},
\end{align}
and
\begin{equation}
\Phi_{appr, 3rd}(x,y) = \mathrm{sign}(z_0) \Phi_1(-x, -y) \left(- i p_x \cos\phi - i p_y \sin\phi + \mathrm{sign}(z_0)  p_z\right)
\end{equation}

\subsection{Electrostatic potential. Fourth quadrant in $\mathbf{q}$ space.  }
\begin{align}
\Phi_{z, 4th} (x, y) & = -\int_{0}^{\infty} dq_x\int_{-\infty}^{0} \, dq_y \frac{ 1}{(2\pi)^2} \,v_c(\mathbf{q}) t_{\mathbf{q}}  e^{i (q_x x + q_y y)} e^{ - |\mathbf{q}| |z_0|}  |\mathbf{q}| \mathrm{sign}(z_0) \notag \\ & =  -\int_{0}^{\infty} dq_x\int_{0}^{\infty} \, dq_y \frac{ 1}{(2\pi)^2} \,v_c(\mathbf{q}) t_{\mathbf{q}}  e^{i ( q_x x - q_y y)} e^{ - |\mathbf{q}| |z_0|}  |\mathbf{q}| \mathrm{sign}(z_0) = \Phi_{z, 1st} (x, -y).
\end{align}

In order to calculate $\Phi_{x, 4th}$ and $\Phi_{y, 4th}$, we should take into account that in the fourth quadrant $q_x = q \cos\phi$, $q_y = - q \sin\phi$. Then
\begin{align}
\Phi_{x, appr, 4th} & =   \frac{ i \cos\phi \, \Phi_{z, 4th, approx}}{\mathrm{sign}(z_0)}, \qquad \Phi_{y, appr, 4th}  = -\frac{i \sin\phi \Phi_{z, 4th, approx} }{\mathrm{sign}(z_0)},
\end{align}
and
\begin{equation}
\Phi_{appr, 4th}(x,y) = \mathrm{sign}(z_0) \Phi_1(x, -y) \left(i p_x \cos\phi - i p_y \sin\phi + \mathrm{sign}(z_0)  p_z\right).
\end{equation}

\subsection{Uni-directional excitation}
\begin{figure}[h!]
	\centering
	\includegraphics[width=3in]{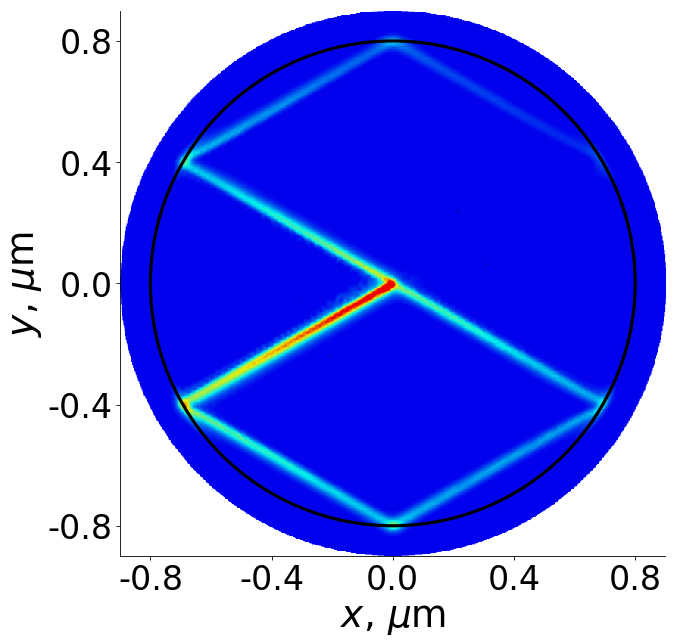}
	\caption{ { Electric field of the plasmons induced in a disk of a hyperbolic material ($\sigma''_x =2.85$ mS, $\sigma''_y =- 0.95$ mS, $\phi = 60^{\circ}$) by an electric dipole $\mathbf{p} = 0.5(\mathbf{e}_x/\cos\phi -\mathbf{e}_y/\sin\phi) - i\mathbf{e}_z$ A$\cdot$m. The disk radius is 800 nm.}}
	\label{fig_three_rays}
\end{figure}
The total potential is
\begin{equation}
\Phi_{appr,tot}(x,y) = \Phi_{appr, 1st}(x,y) + \Phi_{appr, 2nd}(x,y) + \Phi_{appr, 3rd}(x,y) + \Phi_{appr, 4th}(x,y).
\end{equation}

Let us assume that the dipole is above the 2D material, i.e., $\mathrm{sign}(z_0) > 0$. If we consider the dipole $\mathbf{p} = \mathbf{e}_x/\cos\phi- i\mathbf{e}_z$, we obtain
\begin{align*}
\Phi_{appr, 1st}(x,y) & = \Phi_{z, appr, 1st}(x, y) \left(i \cos\phi/\cos\phi -  i\right) = 0, \\
\Phi_{appr, 2nd}(x,y) & =   \Phi_{z, appr, 1st}(-x, y) \left(- i  \cos\phi/\cos\phi - i \right) = -2 i  \Phi_{z, appr, 1st}(-x, y) , \\
\Phi_{appr, 3rd}(x,y) & = \Phi_{z, appr, 1st}(-x, -y) \left(- i \cos\phi/\cos\phi -  i\right) = - 2 i  \Phi_{z, appr, 1st}(-x, -y) , \\
\Phi_{appr, 4th}(x,y) & = \Phi_{z, appr, 1st}(x, -y) \left(i  \cos\phi/\cos\phi -i \right)= 0.
\end{align*}
Thus, such dipoles only excite $\Phi_{appr, 2nd}(x,y)$ and $\Phi_{appr, 3rd}(x,y)$, which describe rays propagating in the direction of the negative $x$ axis. Although, we can only suppress two rays at the same time, by the proper choice of the dipole we can choose which beams will be suppressed. For example, the dipole $\mathbf{p} = \mathbf{e}_x/\cos\phi + i\mathbf{e}_z$ suppresses rays propagating in the negative $x$ direction. On the other hand the dipole $\mathbf{p} = \mathbf{e}_y/\sin\phi + i\mathbf{e}_z$ suppresses rays propagating in the positive $y$ direction. {Finally, the linear polarized dipole,  $\mathbf{p} = \mathbf{e}_x/\cos \phi + \mathbf{e}_y/\sin\phi$, suppresses the rays propagating in the first and third quadrants of the real space. We can also silence only one ray, while allowing for an excitation of the other three, if the dipole is polarized as $\mathbf{p} = 0.5(\mathbf{e}_x/\cos\phi -\mathbf{e}_y/\sin\phi) - i\mathbf{e}_z$ A$\cdot$m (see Fig. \ref{fig_three_rays}). }

\section{An electric dipole in the corner of a rectangle of a hyperbolic material}

{The further control of the uni-directional excitation of the surface plasmons is possible by placing an electric dipole near edges of more complicated shapes. For example, the case of the dipole placed in the upper right corner of a rectangle of the hyperbolic material is presented in Fig. \ref{fig_corner}. One can see that by changing the dipole polarization we can significantly alter the energy deposed into the hyperbolic ray. In fact, the energy flowing through the detector $I_1$ is more than 10 times higher when the dipole polarization is $\mathbf{p} = 2\mathbf{e}_x + 1.15\mathbf{e}_y$ A$\cdot$m (Fig. \ref{fig_corner}a) rather than $\mathbf{p} = 2\mathbf{e}_x - 1.15\mathbf{e}_y$ A$\cdot$m (Fig. \ref{fig_corner}b).}

\begin{figure}[h!]
	\centering
	\includegraphics[width=\linewidth]{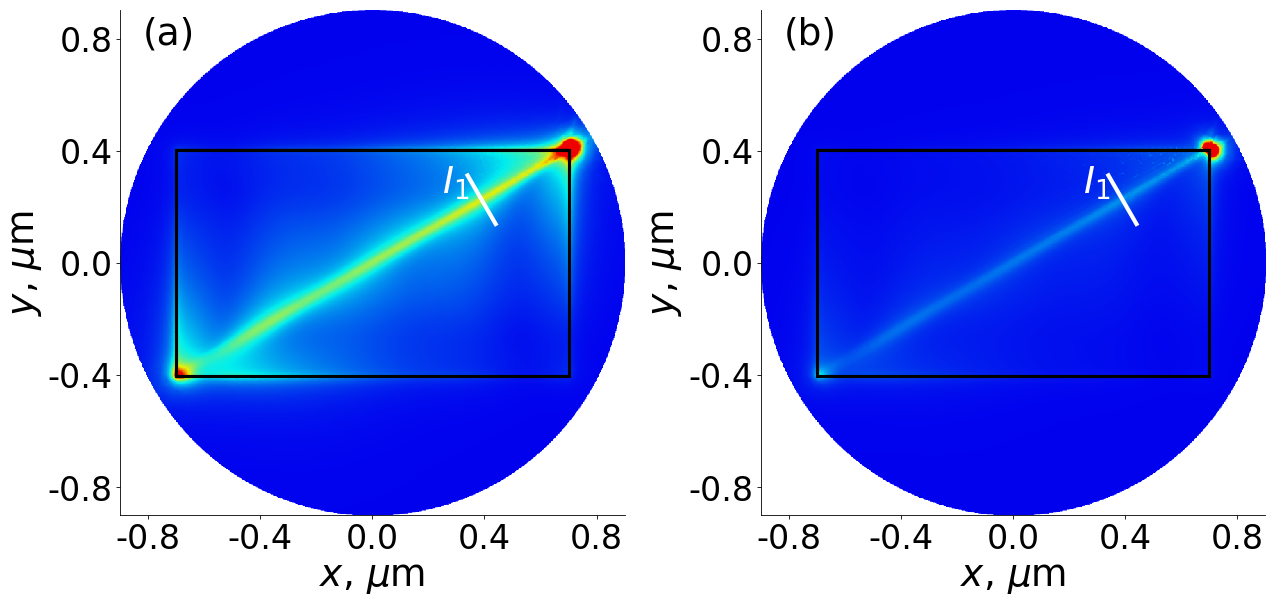}
	\caption{{Spatial distribution of the electric field, $\left|\mathbf{E}\right|$, of the plasmons excited in a rectangle of the hyperbolic material ($\sigma''_x =2.85$ mS, $\sigma''_y =- 0.95$ mS, $\phi = 60^{\circ}$) by an electric dipole placed in the upper right corner of the rectangle 5 nm above the surface. (a) $\mathbf{p} = 2\mathbf{e}_x + 1.15\mathbf{e}_y$ A$\cdot$m, (b) $\mathbf{p} = 2\mathbf{e}_x - 1.15\mathbf{e}_y$ A$\cdot$m.}  }
	\label{fig_corner}
\end{figure}

\section{Uni-directional excitation of surface plasmons in anisotropic material.}
\begin{figure}[h!]
	\centering
	\includegraphics[width=\linewidth]{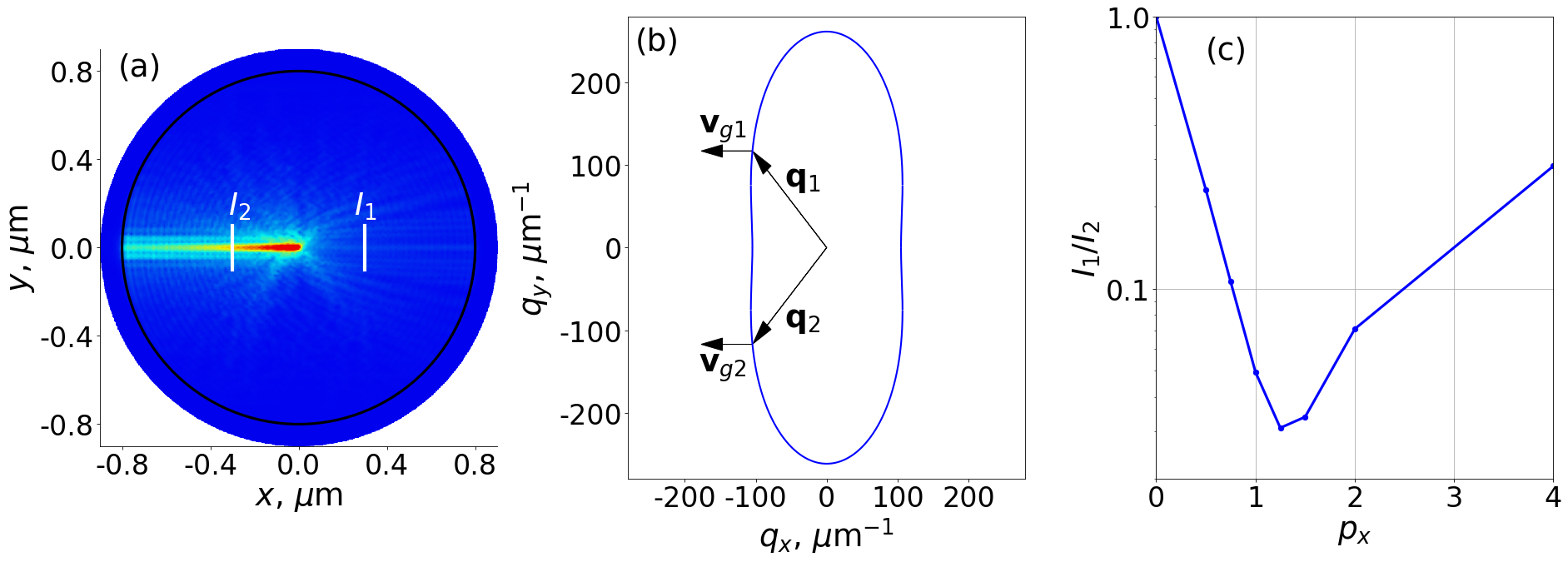}
	\caption{{(a) Spatial distribution of the electric field, $\left|\mathbf{E}\right|$, of the plasmons excited in a disk of the anisotropic material ($\sigma''_{x} = 4.87$ mS, $\sigma''_{y} = 1.95$ mS) by an electric dipole, $\mathbf{p} = \mathbf{e}_x - i\mathbf{e}_z$ A$\cdot$m, placed in a center of the disk 5 nm above the surface. (b) $k$ surface, $\omega(q_x, q_y) = \mathrm{const}$, for plasmons in an anisotropic material. (c) Ratio of intensities, $I_1/I_2$, carried by plasmons through detectors (white lines in panel (a)).}  }
	\label{fig_anis}
\end{figure}

The efficient launching of the uni-directional tightly confined plasmonic rays is possible in a highly anisotropic material that is not hyperbolic, i.e. $\sigma''_{x} \sigma''_{y} > 0$, as can be seen in Fig. \ref{fig_anis}(a). In the anisotropic material, the $k$ surface for the plasmons resembles a highly elongated ellipse with the group velocities predominantly pointing along the optical axis of the material (see Fig. \ref{fig_anis}(b)). This leads to the plasmons carrying their energy in the form of the narrow rays along the optical axis of the material (Fig. \ref{fig_anis}(a)). By using the circular polarized dipole, $\mathbf{p} = p_x \mathbf{e}_x - i \mathbf{e}_z$ A$\cdot$m, we can achieve the efficient uni-directional launching of a single plasmonic ray in such a material (Figs. \ref{fig_anis}(b,c)). {Despite the fact that we don't need an edge to launch the uni-directional plasmons in an anisotropic material, the anisotropic material are disadvantageous compared to the hyperbolic ones. Particular, the plasmonic rays in the anisotropic material can only travel along the directions of the material optical axis. On the other hand, in the hyperbolic material the rays directions of is defined by the imaginary parts of components of the material conductivity tensor and is thus can be controlled by applying an electric bias}.
\end{widetext}

\bibliography{One_way_plasmon_bib}

\end{document}